\documentstyle[preprint,tighten,aps]{revtex}

\newcommand{\act}{{\cal S}}
\newcommand{\be}{\begin{equation}}
\newcommand{\ee}{\end{equation}}
\newcommand{\nl}{\nonumber \\}
\newcommand{\mone}{M_1^{(0)}}
\newcommand{\monesub}{M_{1,sub}^{(1)}}
\newcommand{\intk}{g^2 \frac{4}{3}\int\frac{d^4k}{(2 \pi)^4}}
\newcommand{\vecp}{\vec{p}}
\newcommand{\hkt}{\hat{k}_0}
\newcommand{\hkj}{\hat{k}_j}

\begin{document}
\preprint{\begin{minipage}{2in}\begin{flushright}
 OHSTPY-HEP-T-99-022 \\ MZ-TH/99-64 \end{flushright}
  \end{minipage}}
\input epsf
\draft
\title{{\bf One-Loop Self Energy and Renormalization of the Speed of Light
       for some Anisotropic Improved Quark Actions}}
\author{Stefan Groote}
\address{Newman Laboratory, Cornell University, Ithaca, NY 14853, USA \\
           and \\
    Institut f\"ur Physik der Johannes-Gutenberg-Universit\"at, \\
  Staudingerweg 7, D-55099 Mainz, Germany
       }
\author{Junko Shigemitsu}
\address{Physics Department, The Ohio State University,
  Columbus, OH 43210, USA}
\maketitle

\begin{abstract}
One-loop corrections to the 
fermion rest mass $M_1$, wave function renormalization 
$Z_2$ and speed of light renormalization $C_0$ are presented for lattice 
actions that combine improved glue with clover or D234 quark actions 
and keep the temporal and spatial lattice spacings, $a_t$ and $a_s$, distinct.
We explore a range of values for the anisotropy parameter
 $\chi \equiv a_s/a_t$ 
 and treat both massive and massless fermions.
\end{abstract}

\section{Introduction}
\label{sec:intro}
Examples of successful employment of anisotropic lattices in lattice QCD 
simulations have been increasing lately. They include extensive studies of 
the glueball spectrum\cite{mikecolin}, investigations of heavy hybrid states 
\cite{manke,ianron} and calculations of quarkonium fine structure
 \cite{cppacs}. 
In most cases one is dealing with large states requiring large spatial 
volumes and also signals that can only be extracted from high statistics data.
Working with highly improved actions on coarse lattices helps with the
large volume and statistics problems, however, a coarse temporal lattice 
spacing means that correlation functions  fall off very 
rapidly.  This last problem can be circumvented by going to an 
anisotropic lattice which allows for a much finer temporal grid.  The 
correlation functions can be sampled much more frequently in a given 
physical time region where the signal is still good. 

Another potential use of anisotropic lattices would be in simulations 
of matrix elements in hadronic states with large momenta.  These 
typically occur in semileptonic decays of heavy hadrons.  Once one goes 
beyond spectrum calculations to matrix elements, one is faced with the 
matching problem between operators in the coarse highly improved 
lattice theory and continuum QCD.  In this article we take a first step 
in accumulating necessary renormalization factors based on perturbation theory.
We carry out the one-loop self energy calculation in several anisotropic 
improved quark actions.  This gives us renormalization of the rest mass 
 $M_1$, the wave function renormalization $Z_2$ and the ``speed of light''
renormalization, $C_0$, a quantity 
which will be defined more precisely in the following sections.  We treat 
both massive and massless quarks.  
Perturbation theory can be used not only in operator matchings, but 
also to fix parameters in the lattice actions.  $C_0$ is an example 
of one such parameter.  

\vspace{.1in}
\noindent
In the next section we introduce the gauge and quark actions considered in 
this article.  
Section 3 describes the general formalism that we employ 
for the self energy calculation with massive fermions.  We follow closely 
 the work of the Fermilab group \cite{barth} which we could 
 straightforwardly extend to anisotropic actions. Section 4 discusses 
specific one-loop contributions for mass, wavefunction and speed of 
light renormalizations.  Our results are tabulated in section 5 for 
various choices of actions, fermion masses and degree of anisotropy.  
Some calculational details are left for appendices, where we describe 
Feynman rules and  IR subtractions in our calculations.

\section{Gauge and Quark Actions}
\label{sec:action}

We work with two classes of gauge actions denoted $\act^I_G$ and $\act^{II}_G$
 \cite{alford,gpltsukuba}, with

\begin{eqnarray}
\act^I_G &=& - \beta \sum_{x,s > s^\prime} \frac{1}{\chi} \left\{
c^G_0 \frac{P_{ss^\prime}}{u_s^4} 
+ c^G_1 \frac{R_{ss^\prime}}{u_s^6} 
+ c^G_1 \frac{R_{s^\prime s}}{u_s^6} \right\} \nl
 & & - \beta \;\;\sum_{x,s} \chi \left\{
c^G_0 \frac{P_{st}}{u_s^2 u_t^2} 
+ c^G_1 \frac{R_{st}}{u_s^4 u_t^2} 
+ c^G_1 \frac{R_{ts}}{u_t^4u_s^2} \right\} 
\end{eqnarray}
and 
\begin{eqnarray}
\act^{II}_G &=& - \beta \sum_{x,s > s^\prime} \frac{1}{\chi} \left\{
\frac{5}{3} \frac{P_{ss^\prime}}{u_s^4} 
- \frac{1}{12} \frac{R_{ss^\prime}}{u_s^6} 
- \frac{1}{12} \frac{R_{s^\prime s}}{u_s^6} \right\} \nl
 & & - \beta \;\; \sum_{x,s} \chi \left\{
\frac{4}{3} \frac{P_{st}}{u_s^2 u_t^2} 
- \frac{1}{12} \frac{R_{st}}{u_s^4u_t^2} \right\} .
\end{eqnarray}
The $x$ sum is over lattice sites and the variable $s$ runs over 
spatial directions.  
 $\beta \equiv 2 N_c / g^2$, $\chi$ is the anisotropy parameter 
\be
\chi = a_s/a_t
\ee
 and 
\begin{eqnarray}
P_{\mu \nu} &=& \frac{1}{N_c}  Real \left( Tr \{ U_\mu(x)
 U_\nu(x+a_\mu) U^\dagger_\mu(x+a_\nu) U^\dagger_\nu(x) \} \right) ,  \\
R_{\mu \nu} &=& \frac{1}{N_c}  Real \left( Tr \{ U_\mu(x)
 U_\mu(x+a_\mu) U_\nu(x+2a_\mu) 
U^\dagger_\mu(x+a_\mu + a_\nu) U^\dagger_\mu(x+a_\nu) U^\dagger_\nu(x) \}
 \right)  .
\end{eqnarray}
$u_s$ and $u_t$ are the tadpole improvement parameters $u_0$ for 
spatial and temporal link variables respectively \cite{lepmac}.

The parameters $c^G_0$ and $c^G_1$ in action $\act^I_G$ are constrained 
to satisfy $ c^G_0 + 8 c^G_1 = 1$.  The Symanzik improved gauge action,
 in which $O(a^2)$ errors are removed, 
corresponds to $c^G_0 = 5/3$ and $c^G_1 = - 1/12$ \cite{weisz}, 
whereas $c^G_0 = 3.648$ 
and $c^G_1 = - 0.331$, for $\chi = 1$,
 leads to one of the RG improved Iwasaki actions 
\cite{iwasaki}.  In $\act^{II}_G$ parameters have been fixed 
to the Symanzik values.  We will be working mainly with Symanzik improved 
actions and present RG improved results only for a few cases. 
We note that the action $\act^{II}_G$ is intrinsically asymmetric
  even for the isotropic limit $\chi = 1$.

\vspace{.1in}
The most highly improved quark action that we have analysed is the 
$D234$ action \cite{alford}.  
\begin{eqnarray}
\label{sd234a}
\act^I_{D234} &=& a_s^3 a_t \sum_x \overline{\Psi}_c 
\left\{  \gamma_t \frac{1}{a_t} (\nabla_t - \frac{1}{6} 
C_{3t} \nabla_t^{(3)}) + 
\frac{C_0}{a_s} \vec{\gamma} \cdot (\vec{\nabla} -
 \frac{1}{6} 
C_{3} \vec\nabla^{(3)}) + m_0 \right. \nl
 &  & - \frac{r a_s}{2} \left [ \frac{1}{a_t^2} ( \nabla_t^{(2)} 
- \frac{1}{12} C_{4t} \nabla_t^{(4)} )
+  \frac{1}{a_s^2} \sum_j ( \nabla_j^{(2)} 
- \frac{1}{12} C_{4} \nabla_j^{(4)} ) \right ] \nl
  & & - r a_s \left. \frac{C_F}{4} 
\frac{i \sigma_{\mu \nu} \tilde{F}^{\mu \nu}} {a_\mu a_\nu} 
\right\} 
\Psi_c    \\
\label{sd234b}
 &=&  \sum_x \overline{\Psi}_L 
\left\{  \gamma_t (\nabla_t - \frac{1}{6} 
C_{3t} \nabla_t^{(3)}) + 
\frac{C_0}{\chi} \vec{\gamma} \cdot (\vec{\nabla} -
 \frac{1}{6} 
C_{3} \vec\nabla^{(3)}) + a_t m_0 \right. \nl
 &  & - \frac{r }{2} \left [ \chi ( \nabla_t^{(2)} 
- \frac{1}{12} C_{4t} \nabla_t^{(4)} )
+  \frac{1}{\chi} \sum_j ( \nabla_j^{(2)} 
- \frac{1}{12} C_{4} \nabla_j^{(4)} ) \right ] \nl
  & & - r \left. \frac{C_F}{4} 
i \sigma_{\mu \nu} \tilde{F}^{\mu \nu} \frac{a_s a_t}{a_\mu a_\nu} 
\right\} 
\Psi_L .
\end{eqnarray}
The quark fields $\Psi_c$ and the dimensionless lattice 
fields $\Psi_L$ are related through
\be
\Psi_L = a_s^{3/2} \Psi_c .
\ee
The dimensionless derivatives $\nabla^{(n)}$ and field strength 
tensors $\tilde{F}^{\mu \nu}$ are tadpole improved \cite{lepmac} and 
defined in the Appendix. We use 
the convention $\sigma_{\mu \nu} = \frac{1}{2}[\gamma_\mu, \gamma_\nu]$ 
and set $r = 1$ in all our calculations.  
At tree-level the coefficients $C_0$, $C_3$, $C_{3t}$, $C_4$, 
$C_{4t}$ and $C_F$ are equal to one.
  The quark action is then tree-level accurate 
through $O(a_s^3)$ and $O(a_t^3)$.  $C_0$ is what we call the ``speed of 
light''.  This parameter is adjusted, in general either perturbatively 
or nonperturbatively, to ensure correct dispersion relations for particles.  
In anticipation of working on anisotropic lattices with $a_t$ much finer 
than $a_s$, one can drop the higher order improvement terms in 
the temporal derivatives by setting $C_{3t} = C_{4t} = 0$, without 
loosing accuracy.  We call this action $\act^{II}_{D234}$.
\begin{eqnarray}
\act^{II}_{D234} &=&  \sum_x \overline{\Psi}_L 
\left\{  \gamma_t \nabla_t +
\frac{C_0}{\chi} \vec{\gamma} \cdot (\vec{\nabla} -
 \frac{1}{6} 
C_{3} \vec\nabla^{(3)}) + a_t m_0 \right. \nl
 &  & - \frac{r }{2} \left [ \chi  \nabla_t^{(2)} 
+  \frac{1}{\chi} \sum_j ( \nabla_j^{(2)} 
- \frac{1}{12} C_{4} \nabla_j^{(4)} ) \right ] \nl
  & & - r \left. \frac{C_F}{4} 
i \sigma_{\mu \nu} \tilde{F}^{\mu \nu} \frac{a_s a_t}{a_\mu a_\nu} 
\right\} 
\Psi_L 
\end{eqnarray}
The familiar $O(a)$ accurate clover quark \cite{SWaction} 
action corresponds to 
setting $C_3=C_4=0$ in the above and using a less improved field strength 
tensor $F^{\mu \nu}$ (also defined in the Appendix) rather than 
$\tilde{F}^{\mu \nu}$. 
\begin{eqnarray}
\act_{clover} &=&  \sum_x \overline{\Psi}_L 
\left\{  \gamma_t \nabla_t +
\frac{C_0}{\chi} \vec{\gamma} \cdot \vec{\nabla} 
 + a_t m_0 \right. \nl
 & & - \frac{r }{2} \left [ \chi  \nabla_t^{(2)} 
+  \frac{1}{\chi} \sum_j  \nabla_j^{(2)} 
 \right ] 
 - r \left. \frac{C_F}{4} 
i \sigma_{\mu \nu} F^{\mu \nu} \frac{a_s a_t}{a_\mu a_\nu} 
\right\} 
\Psi_L
\end{eqnarray}

\vspace{.1in}
We have carried out one-loop self energy calculations for several combinations 
of the above gauge and quark actions, for both massless and massive quarks. 
We list the specific actions considered in Table I.  For actions $\act^A$ and 
$\act^{A^\prime}$  massless results have already appeared in \cite{aoki}. 
We agree with their results and we include these cases here for completeness.
 With action $\act^C$ 
we treat only the massless case, since our formalism for massive quarks, 
following \cite{barth}, requires that the only time derivatives be 
in the $\nabla_t$ and $\nabla_t^{(2)}$ terms. 
Both $\act^{II}_{D234}$ and $\act_{clover}$ satisfy this condition, but 
$\act^I_{D234}$ does not.

\section{General Formalism for Self Energy Calculations}
\label{sec:formalism}

In this section we summarize the formalism for self energy calculations, 
along the lines of reference \cite{barth}. Perturbative 
calculations for massive Wilson quarks are also described in reference 
\cite{kura}. 
 We concentrate 
on the massive case, since massless lattice perturbation theory has been 
in the literature for decades.

\vspace{.1in}
For massive fermions we use either $\act^{II}_{D234}$ or $\act_{clover}$.
The fermion self energy $\Sigma(p)$ is defined in terms of the momentum 
space propagators $\overline{G}(p)$ and $\overline{G}_0(p)$ for the 
full and free theories respectively, as 
\be
\label{ginv}
\overline{G}^{-1}(p) = \overline{G}^{-1}_0(p) - \Sigma(p) .
\ee
Carrying out the Fourier transform in $p_0$ one defines
\begin{eqnarray}
\label{def}
G(t,\vecp) & = & \int^{\pi/a_t}_{-\pi/a_t} \frac{dp_0}{2 \pi} 
e^{ip_0 t} \overline{G}(p_0,\vecp)  \nl
  & \equiv & {\cal Z}_2(\vecp) e^{-E(\vecp)t} \Gamma_{proj} \;
+ \; \ldots  \;\; .
\end{eqnarray}
$\Gamma_{proj}$ is a projection operator in Dirac space.  The ellipses 
refer to lattice artifacts and additional multi-particle states that 
could be created by the lattice fermion field operator $\Psi$ beyond the 
single quark state.  The rest mass, $M_1$, is defined as
\be
M_1 = E(\vecp = \vec{0}) .
\ee
We do not consider the kinetic mass, $M_2$ \cite{barth} in this article. 
We will renormalize at the point $(p_0, \vecp) = (i M_1, \vec{0})$ and 
define the wave function renormalization constant 
\be
Z_2 = {\cal Z}_2(\vecp = \vec{0}).
\ee
For a zero spatial momentum quark propagating forward in time one 
 expects ($t > 0$)
\begin{eqnarray}
\label{def0}
G(t,0) & = & \int^{\pi/a_t}_{-\pi/a_t} \frac{dp_0}{2 \pi} 
e^{ip_0 t} \overline{G}(p_0,0)  \nl
  & \equiv &  Z_2 e^{-M_1t} \frac{1 + \gamma_0}{2} 
 \; + \; \ldots \;\; .
\end{eqnarray}
Our  goal in this section is to relate $Z_2$ and $M_1$ to parameters in the 
action and to $\Sigma(p)$.  In order to orient ourselves, however,
 it is useful to first consider the free case with $\Sigma(p) = 0$.

\subsection{Free Anisotropic Propagator}
The free propagator $\overline{G}_0(p_0,\vecp=0)$
 for both actions $\act^{II}_{D234}$ 
and $\act_{clover}$ becomes (for $r = 1$)
\begin{eqnarray}
\frac{1}{a_t} \, \overline{G}_0(p_0,\vecp = 0) &=&
 \frac{1}{i\gamma_0 \sin(a_tp_0) + a_tm_0 + \chi 
 - \chi \cos(a_tp_0) }  \nl
 &=& {{-i\gamma_0 \sin(a_tp_0) + a_tm_0 + \chi 
 - \chi \cos(a_tp_0)  } \over {(\sin(a_tp_0))^2 + [(a_tm_0 +\chi) -
 \chi \cos(a_tp_0)]^2 }}.
\end{eqnarray}
In terms of the variable
\be 
z \equiv e^{i a_t p_0} = e^{-a_tE}
\ee
($ p_0 = iE$), one finds two zeros of the denominator corresponding to 
positive energy solutions.
\be
\label{m1}
z_1 = {{(a_tm_0 + \chi) - \sqrt{(a_tm_0 +\chi)^2 + 1 - \chi^2 }} \over {\chi - 1}}
\ee
and
\be
\tilde{z}_1 = {{(a_tm_0 + \chi) - \sqrt{(a_tm_0+\chi)^2 + 1 - 
\chi^2 }} \over {\chi + 1}}.
\ee
The other two zeros, $z_2$ and $\tilde{z}_2$ correspond to negative 
energy solutions, $z_2 = 1/z_1$ and $ \tilde{z}_2 = 1/\tilde{z}_1$.  
The integral over $p_0$ in (\ref{def0}) can be done as a contour 
integral around the unit circle in the variable $z$.  One picks up 
contributions from both positive energy solutions (for $t > 0$).
\begin{eqnarray}
pole\;\; at\;\; z_1 & : &\qquad \frac{(1 + \gamma_0)}{2}
 \frac{e^{-\mone t}}{
\sqrt{(a_tm_0 + \chi)^2 + 1 - \chi^2}} \label{pole} ,\\
pole\;\; at\;\; \tilde{z}_1 & : &\qquad
 \frac{(1 - \gamma_0)}{2} \frac{e^{-\tilde{M}^{(0)}_1t}}{
\sqrt{(a_tm_0 + \chi)^2 + 1 - \chi^2}}  ,
\end{eqnarray}
with
\be
a_t\mone = - \ln(z_1), \qquad \qquad \qquad a_t\tilde{M}^{(0)}_1 =
 - \ln(\tilde{z}_1).
\ee
Clearly, $z_1$ is the physical positive energy solution. 
The second solution $\tilde{z}_1$ is a lattice artifact, similar to 
the time doubler for $r \neq 1$ in isotropic actions.  
The solution $\tilde{z}_1$ disappears in the isotropic limit, 
$\chi \rightarrow 1$, where 
$a_t\tilde{M}^{(0)}_1 \rightarrow \infty$.  
In the same limit the physical solution 
$z_1$ goes over into the well known result 
\be
z_1 \rightarrow \frac{1}{1 + a_tm_0}.
\ee
The gap between $\tilde{M}^{(0)}_1$ and $\mone$, measured in units of $1/a_s$ 
is
\be
a_s (\tilde{M}^{(0)}_1 - \mone ) = \chi \; \ln\frac{(\chi + 1)}{(\chi - 1)},
\ee
independent of $m_0$.  This becomes $\infty$ at $\chi = 1$ and 
approaches $2$ as $\chi \rightarrow \infty$.  The size of this gap, 
$a_s \delta E \geq 2$, is hence equal to or larger
 than the amount by which conventional spatial doublers
 are raised through the Wilson mechanism.  We will henceforth 
ignore $\tilde{z}_1$ and concentrate 
on the physical pole at $z = z_1$.
Comparing (\ref{pole}) with (\ref{def0}) one sees that there is nontrivial 
mass dependent wave function renormalization even at tree-level with 
\be
\label{z20}
Z_2^{(0)} = \frac{1}{\sqrt{(a_tm_0 + \chi)^2 + 1 - \chi^2}} =
\frac{1}{\chi \; \sinh(a_t\mone) + \cosh(a_t\mone)}.
\ee
This has been pointed out several times in the literature \cite{luesch,fermi}.

A useful way to rewrite (\ref{m1}) is 
\be
\label{ampchi}
a_tm_0 + \chi = \chi \; \cosh(a_t \mone) + \sinh(a_t\mone).
\ee

\subsection{Mass Renormalization}

\noindent
In the interacting case one has a nontrivial $\Sigma(p)$ which 
we write as

\vspace{.1in}
\be
\label{selfe}
a_t\Sigma(p) = i \gamma_0 B_0(p,m_0) \sin(a_tp_0) + 
 i\frac{1}{\chi} \sum_j [\gamma_j  B_j (p,m_0) \sin(a_s p_j) ]
+ C(p,m_0).
\ee
The $\vec{p} = 0$ propagator becomes
\be
\label{gpfull}
\frac{1}{a_t} \, \overline{G}(p_0,\vecp = 0) 
 = {{-i\gamma_0 (1-B_0) \sin(a_tp_0) + a_tm_0 + \chi - C 
 - \chi \cos(a_tp_0)  } \over {(1-B_0)^2[\sin(a_tp_0)]^2 + [(a_tm_0+\chi) -
 C - \chi \cos(a_tp_0)]^2 }} ,
\ee
where, $B_0=B_0(p_0,m_0)$ and $C=C(p_0,m_0)$ are evaluated at $\vec{p} = 0$.  
If $p_0 = i E$ is the location of a pole in (\ref{gpfull}), the following 
implicit equation must be satisfied.
\be
\label{implicit}
(1-B_0(iE,m_0)) \sinh(a_t E) = \pm [ (a_t m_0 + \chi) - C(iE,m_0
) - \chi \cosh(a_t E) ].
\ee
One can check that the ``+'' sign leads to the pole $\mone$ in the 
free limit.  Hence, the implicit equation for $M_1$ is given by
\be
\label{m1full}
 \chi \; \cosh(a_t M_1) + \sinh(a_t M_1) = a_tm_0 + \chi + B_0(iM_1,m_0) 
\sinh(a_t M_1) - C(iM_1,m_0).
\ee

\vspace{.1in}
\noindent
In a perturbative calculation of $M_1$ one expands
\be
M_1 = \mone + \alpha_s M_1^{(1)} + O(\alpha_s^2).
\ee
$B_0$ and $C$ in (\ref{m1full}) start out $O(\alpha_s)$, so through 
one-loop their argument can be replaced by the tree-level $\mone$. Expanding 
the LHS also through $O(\alpha_s)$ and taking (\ref{ampchi}) into account, one 
finds
\begin{eqnarray}
\label{m11}
 \alpha_s a_t M_1^{(1)}
 &=&  \frac{B_0(i\mone,m_0) \sinh(a_t \mone) - C(i\mone,m_0)} 
 { \chi \sinh(a_t \mone) + \cosh(a_t \mone) }  \nl
 &=&  - \; Z_2^{(0)}  \; tr \,\left\{\frac{(\gamma_0 + 1)}{4}\, a_t \,\Sigma
(p_0 = i \mone, \vec{p} = 0) \right \} ,
\end{eqnarray}
where the trace is taken over Dirac space.
We note that in the $ \mone = 0, m_0=0$ limit, the $\gamma_0$ part of the 
trace $tr \{(\gamma_0 + 1) \Sigma \} $ does not contribute and one has 
\be
 \alpha_s a_t M_1^{(1)}(0)
 =  - \; tr \left\{ a_t \, \Sigma(0)  \right \} / 4 = - C(0,0).
\ee
In order to have massless quarks remain massless under renormalization, 
one needs to carry out additive mass renormalization and $M_1^{(1)}$ in 
(\ref{m11}) requires a subtraction. This subtraction must be done
 without jeopardizing the pole condition (\ref{m1full}). 
There is a  standard way to accomplish this.
Let 
 $m_c$ be the value of the bare quark mass parameter $m_0$ for which 
the physical quark rest mass vanishes ($M_1 = 0)$.  Eqn.(\ref{m1full}) then 
tells us that $m_c$ is implicitly defined through
\be
\label{mc}
a_t m_c - C(0,m_c) = 0 .
\ee
In equations such as (\ref{gpfull}) or (\ref{m1full}) one always has the
 combination $a_t m_0 - C$.  Using (\ref{mc})
 one can add and subtract $a_t m_c$ so that
\be
a_t m_0- C \rightarrow a_t (m_0 - m_c) - (C - C(0,m_c)) = a_t m - \tilde{C}.
\ee
Previous derivations go through with $m_0$ replaced by
\be 
m \equiv m_0 - m_c
\ee
and $C(iM_1,m_0)$ by
\be
\tilde{C}(iM_1,m_0) = C(iM_1,m_0) - C(0,m_c).
\ee
In most lattice simulations, $m_c$ and hence also
$m$ are determined nonperturbatively from 
the simulations themselves.  For $\tilde{C}$, however, one still often uses 
the one-loop result
\be
\label{pertct}
\tilde{C}(i\mone,m_0) = \tilde{C}(i\mone,m) = C(i\mone,m) - C(0,0).
\ee
$\mone$ is now given in terms of $m$ rather than $m_0$.  
We will be presenting our results as functions of $a_s\mone$, with 
the understanding that the shift $m_0 \rightarrow m_0 - m_c$ has been 
carried out and that, for instance, $\mone$ is given by
\be
a_tm + \chi = \chi \; \cosh(a_t \mone) + \sinh(a_t\mone)
\ee
rather than by (\ref{ampchi}).
In (\ref{m11}) one needs to replace $C$ by $\tilde{C}$.  Our final formula
for the one-loop mass correction, measured in units of $1/a_s$ then 
becomes
\begin{eqnarray}
\label{m11b}
 \alpha_s a_s M_{1,sub}^{(1)}
 &=&  \chi \;\frac{B_0(i\mone,m) \sinh(a_t \mone) - \tilde{C}(i\mone,m)} 
 { \chi \sinh(a_t \mone) + \cosh(a_t \mone) }  \nl
 &=&  -  \; Z_2^{(0)}
  \; tr \left\{\frac{(\gamma_0 + 1)}{4}\,\left[ a_s \,\Sigma
(p_0 = i \mone, \vec{p} = 0,m) - a_s \Sigma(0,\vec{0},0) \right]\right \} .
\end{eqnarray}
This expression vanishes automatically for $\mone = m = 0$.
We prefer to measure dimensionful quantities in terms of $1/a_s$ rather than 
$1/a_t$.  When exploring $\chi \geq 1$ it makes more sense to 
fix $a_s$ and let $a_t$ be arbitrarily fine, rather than to fix $a_t$ 
and let $a_s$ become arbitrarily coarse.  
In the isotropic limit (\ref{m11b}) agrees with formulas in the 
literature \cite{barth,kura}.

\subsection{Wave Function Renormalization}
In order to extract a general formula for the wave function renormalization 
$Z_2$ we need to find the residue of $\overline{G}(p_0,\vecp = 0)$ at the 
pole $p_0 = iM_1$.  In terms of the variable
 $z$ the Fourier transform in (\ref{def0}) has the form
\be
\oint_{|z|=1} \frac{dz}{(2 \pi i) z} (z)^{t/a_t} \frac{g(z)}{f(z)} ,
\ee
where the integral is taken over the unit circle.  To find the residue we 
expand the denominator around $z_1 = e^{-a_t M_1}$ 
\be
f(z) = (z-z_1) \left(\frac{d \,f}{dz} \right)_{z=z_1} \;\; + \; \; 
\ldots \;\; .
\ee
The contribution from the physical pole to $G(t,0)$ is then
\be
e^{- M_1 t} \left( \frac{g(z)}{z\, f^\prime(z)} \right)_{z = z_1}.
\ee
One finds for the numerator
\be
g(z=z_1) = (\gamma_0 + 1) ( 1 - B_0(iM_1,m) ) \sinh(a_t M_1)
\ee
and for the denominator
\begin{eqnarray}
\label{denom}
& &2 \, (1-B_0(iM_1,m)) \sinh(a_tM_1) \times  \nl
 & & \left\{ \chi \sinh(a_tM_1) + \cosh(a_tM_1)  
+ \left(i \frac{d}{d(a_tp_0)} [ i B_0(p_0,m) \sin(a_tp_0) + C(p_0,m)]
 \right)_{p_0 = iM_1} \right\}
\end{eqnarray}
using
\be
 \left(z \frac{d \,f}{dz} \right)_{z = z_1} = -i \left( \frac{d \,f}{d(a_tp_0)}
\right)_{p_0 = iM_1}.
\ee
One can now read off $Z_2$ and after recognizing the last term in 
(\ref{denom}) as derivatives acting on different parts of 
$a_t \Sigma(p_0,\vecp=0,m)$, one 
obtains
\be
\label{z2inv}
Z_2^{-1}
=  \chi \sinh(a_tM_1) + \cosh(a_tM_1)  
+ i\, tr \left( \frac{(\gamma_0 + 1)}{4}
 \frac{d}{dp_0} \Sigma(p_0,\vecp=0,m)
 \right)_{p_0 = iM_1} .
\ee
The one-loop approximation to $Z_2$ is obtained by expanding $M_1$ once 
again in $\alpha_s$. 
\begin{eqnarray}
\label{z2inv1}
Z_2^{-1}
&=&  \chi \sinh(a_t\mone) + \cosh(a_t\mone)  
 + \alpha_s a_t M_{1,sub}^{(1)}\; (\chi \cosh(a_t\mone) + 
\sinh(a_t\mone) ) \nl
& & + i\, tr \left( \frac{(\gamma_0 + 1)}{4}
 \frac{d}{dp_0} \Sigma(p_0,\vecp=0,m)
 \right)_{p_0 = i\mone}   \nl
&=& Z_2^{(0)-1} [\; 1 
 + \frac{\alpha_s}{\chi} a_s M_{1,sub}^{(1)} \;(\chi \cosh(a_t\mone) + 
\sinh(a_t\mone)) \, Z_2^{(0)}  \nl
& & + i\, tr \left( \frac{(\gamma_0 + 1)}{4}
 \frac{d}{dp_0} \Sigma(p_0,\vecp=0,m)
 \right)_{p_0 = i\mone}  
Z_2^{(0)} \, ] \qquad + \qquad O(\alpha_s^2) .
\end{eqnarray}
In the last expression we have found it convenient to factor out 
the tree-level  $Z_2^{(0)-1}$.  Equations (\ref{z2inv}) and (\ref{z2inv1})
go over into the formulas of \cite{barth} in the isotropic limit.

\subsection{Speed of Light Renormalization}
In order to discuss renormalization of the speed of light
 one needs to look at the inverse momentum space 
propagator at small but nonzero spatial momentum.
\begin{eqnarray}
a_t\overline{G}^{-1}(p)  & = & a_t\overline{G}_0^{-1}(p) - a_t \Sigma(p) \nl
 & = & i \, \gamma_0(1 - B_0) \sin(a_tp_0) + i \,  \frac{1}{\chi} \sum_j [
\gamma_j \, (C_0K_j - B_j) \sin(a_sp_j)]   \nl
 & & + a_tm + \chi - \chi \cos(a_t p_0) - C ,
\end{eqnarray}
with $K_j = 1$ for $\act_{clover}$ and $K_j = (4 - \cos(a_sp_j))/3$ for 
$\act^{I,II}_{D234}$.  One can rewrite $\overline{G}^{-1}(p)$ as
\begin{eqnarray}
a_t\overline{G}^{-1}(p)  & = & (1-B_0) \left\{
i \, \gamma_0 \,\sin(a_tp_0) + i \,  \frac{1}{\chi} \sum_j [
\gamma_j\frac{(C_0K_j - B_j)}{(1-B_0)} \sin(a_sp_j)] \right\} \nl
 & & + a_tm + \chi - \chi \cos(a_t p_0) - C .
\end{eqnarray}
 $C_0$ is adjusted so that for small $a_sp_j$ the relative coefficient
of the $\gamma_0 \sin(a_tp_0)$  and the $\gamma_j a_sp_j/\chi$
 terms 
remains equal to unity.  
$(C_0K_j-B_j) \sin(a_sp_j)=(C_0 - B_j)a_s p_j$ for all quark actions 
in this limit 
(of course, $K_j \sin(a_sp_j)$ 
is a better approximation to the continuum $a_sp_j$ in the $D234$ action 
than in the clover action), and one has
\be
\label{c0}
C_0 = 1 + B_j(C_0) - B_0(C_0)  \; \approx \; 
 1 + B_j(C_0=1) - B_0(C_0=1) \quad + \quad O(\alpha_s^2) .
\ee
Just as with $Z_2$ we will define the speed of light renormalization at 
the zero spatial momentum mass shell point $ p = (iM_1,\vec{0})$.  
From (\ref{selfe}) the two terms $B_j$ and $B_0$ needed for 
$C_0$ at one-loop  can be extracted through
\begin{eqnarray}
\label{bj}
B_j &=& -i \frac{\chi}{4} tr \left( \gamma_j \, \frac{\partial}
{\partial(a_s p_j)}
 a_t \Sigma(p) \right)_{p = (i\mone,\vec{0})},  \\
\label{b00}
B_0 &=& -\frac{1}{4}  \frac{ tr (\gamma_0 \, a_t\Sigma(i\mone,\vec{0}) ) }
{\sinh(a_t\mone)} \quad \qquad m > 0  \\
 &or&  \nl
\label{b0}
 &=& - \frac{i}{4} tr \left( \gamma_0 \, \frac{\partial}{\partial p_0}
 \Sigma(p) \right)_{p = (0,\vec{0})} \qquad m=0 .
\end{eqnarray}
We note that in the massive case there is nontrivial renormalization of $C_0$
even in the isotropic limit $\chi = 1$, due to 
 our noncovariant mass shell condition, $ p = (iM_1,\vec{0})$.
  Nevertheless, we believe the above definition of 
the renormalization of the speed of light is a sensible and 
physical one. 

\section{One-Loop Contributions to $\Sigma(p)$}
\label{sec:diagrams}
In the previous section one-loop corrections for $M_1$, $Z_2$ and $C_0$ 
were determined in terms of traces over $\Sigma(p)$ or over derivatives 
acting on $\Sigma(p)$.  In this section we describe the lattice 
perturbation theory diagrams that contribute to $\Sigma(p)$ at one loop. 
For all quark actions considered one can write
\be
\Sigma(p) = \Sigma^{reg}(p) + \Sigma^{tad}(p) + \Sigma^{t.i.}(p).
\ee
$\Sigma^{reg}$ is the regular ``rainbow'' diagram, the only diagram that exists
in a continuum self energy calculation. $\Sigma^{tad}$ denotes contributions 
from the lattice artifact tadpole diagram and $\Sigma^{t.i.}$ comes from 
perturbatively expanding the $u_s$'s and $u_t$'s entering definitions of 
tadpole improved derivatives (see Appendix).  The main idea behind 
tadpole-improved perturbation theory \cite{lepmac} is to have 
$\Sigma^{t.i.}$ cancel the bulk of $\Sigma^{tad}$.  We find in many 
instances, especially with $\act_{clover}$, 
 that cancellation is complete if one uses the Landau link definition 
for $u_s$ and $u_t$ and works in Landau gauge.

\subsection{$\Sigma^{reg}(p)$}
In terms of the gauge propagator $D_{\mu\nu}$, quark propagator
 $\overline{G}_0$ and the vertex functions $V_\mu$,
one can write $a_t\Sigma^{reg}(p)$ as the  
following loop integral over the dimensionless momentum variables
$-\pi \leq k_\mu \leq \pi$.
\begin{eqnarray}
\label{sigreg}
& & a_t \Sigma^{reg}(p)  \nl
& & = g^2 \frac{4}{3}\sum_{\mu,\nu} \int \frac{d^4k}
{(2 \pi)^4 }
\left\{ V_\mu(ap,ap-k)\; \frac{\overline{G}_0(ap-k)}{a_t} \;
V_\nu(ap-k,ap) \right\} D_{\mu\nu}(k,\alpha_g)  \nl
&& = g^2 \frac{4}{3}\sum_{\mu,\nu} \int \frac{d^4k}
{(2 \pi)^4 }
\left\{ V_\mu(ap,ap-k) [ -i  \gamma \cdot K 
\sin + \Omega ]_{(ap-k)} V_\nu(ap-k,ap) \right\} \frac{D_{\mu\nu}
(k,\alpha_g)}{(K^2 \sin^2+ \Omega^2)_{(ap-k)}} \nl
&&
\end{eqnarray}
where, $ap$ stands for $(a_tp_0,a_s\vecp)$ and $K \sin $ for $ K_0 
\sin((ap-k)_0)$ or for $ K_j \sin((ap-k)_j)/\chi$.  
  $K_\mu(k)$, $\Omega(k)$, 
$V_\mu(k^\prime,k)$ and $D_{\mu\nu}(k,\alpha_g)$ are detailed in the
 Appendix.  The argument $\alpha_g$ in the gauge propagator comes from 
the gauge fixing term, 
with $\alpha_g =1$ and $\alpha_g=0$ corresponding to 
Feynman and Landau gauges respectively.  
Our codes have been written for general $\alpha_g$ and we have used 
gauge invariance of $M_1$ and $C_0$ as one check on our results.

\vspace{.1in}
\noindent
Equation (\ref{sigreg}) has the familiar form for a self energy integral. 
The only subtlety is to verify that one is indeed calculating $\Sigma^{reg}$ 
measured in units of $1/a_t$, given  the conventions in our Feynman rules. 
As explained in the Appendix, we choose to work with a dimensionless
momentum space  gauge propagator 
$D_{\mu\nu}$.  It comes from the Fourier transform of the 
dimensionless correlator 
$\langle (a_\mu A_\mu) \; (a_\nu A_\nu) \rangle$.  The relation between
$D_{\mu \nu}$ and a more conventional  propagator of dimension
1/(energy)$^2$, denoted $\tilde{D}_{\mu\nu}$, is
\be
\label{dtilde}
D_{\mu\nu} = \frac{a_\mu a_\nu}{a_s^3 a_t} \, \tilde{D}_{\mu\nu} .
\ee
Our vertex functions, $V_\mu$, are also subtlely different from those 
in isotropic lattice perturbation theory.  They keep tract of the $1/a_\mu$ 
in the derivatives, i.e. of whether one has a $1/a_s$ or $1/a_t$ there. 
If $\tilde{V}_\mu$ are vertex functions normalized such that
$\tilde{V}_\mu  \rightarrow - i \gamma_\mu$ for all $\mu$ 
in the continuum limit, then the relation to the $V_\mu$ of (\ref{sigreg})
is given by
\be
\label{vtilde}
V_\mu = \frac{a_t}{a_\mu} \tilde{V}_\mu .
\ee
Using (\ref{dtilde}) and (\ref{vtilde}) one can write
\begin{eqnarray}
\label{sigreg2}
 a_t \Sigma^{reg}(p)  
 & = &g^2 \frac{4}{3}\sum_{\mu,\nu} \int \frac{d^4k}
{(2 \pi)^4 }
\left\{ \left(\frac{a_t}{a_\mu}\tilde{V}_\mu \right)\; 
\frac{\overline{G}_0}{a_t} \;
\left(\frac{a_t}{a_\nu}\tilde{V}_\nu \right) \right\} 
\left( \frac{a_\mu a_\nu}{a_s^3a_t}  \tilde{D}_{\mu\nu} \right) \nl
 & = &a_t\, g^2 \frac{4}{3}\sum_{\mu,\nu} \int \frac{d^4k}
{(2 \pi)^4 a_s^3 a_t}
\left\{ \tilde{V}_\mu \; 
\overline{G}_0 \;
\tilde{V}_\nu  \right\} 
 \tilde{D}_{\mu\nu}  .
\end{eqnarray}

\vspace{.1in}
\noindent
To evaluate (\ref{sigreg}) we made extensive use of the symbolic 
manipulation package Mathematica.  The integrals themselves were done using 
the VEGAS program \cite{vegas}. 
 The various steps involving Mathematica were to 
1. calculate the products $V_\mu [-i \gamma \cdot K \sin + \Omega] V_\nu$; 
2. carry out the Dirac traces such as $tr \{(1 + \gamma_0) \Sigma \}$ ;
3. take derivatives with respect to external momenta ;
4. put things on the mass shell $p = (iM_1,\vec{0})$ ;
 and 5. use trigonometric identities to re-express the full integrands
in (\ref{sigreg}) in terms 
of  powers of $\hat{k}_\mu \equiv 2 \sin(k_\mu/2)$.
The last step facilitated speedy evaluation of the integrand by 
VEGAS.

\vspace{.1in}
\noindent
Both $M_1$ and $C_0$ are physical quantities. In addition to being gauge 
invariant they are also IR finite.  The wave function renormalization $Z_2$,
on the other hand, is gauge dependent and also generally 
logarithmically IR divergent.  In any calculation of a physical 
quantity this IR divergence will eventually
 be cancelled by vertex corrections and/or 
matching to continuum operators.
In this article we will isolate the gauge dependent 
IR divergence in $Z_2$, verify that 
it is the same as in the corresponding continuum theory and present 
results for the remaining IR finite parts.  The IR divergence is 
found in the contribution from $\Sigma^{reg}$ to $Z_2$.  
More specifically it resides in the following term in (\ref{z2inv1}) 
\be
\label{z2ir}
 i\, tr \left( \frac{(\gamma_0 + 1)}{4}
 \frac{d}{dp_0} \Sigma^{reg}(p_0,\vec{p}=0,m)
 \right)_{p_0 = i\mone}  
Z_2^{(0)} .
\ee
We adopt the method of reference \cite{kura} to subtract
IR divergent contributions inside integrands and rewrite (\ref{z2ir}) as
\begin{eqnarray}
\label{z2sub}
 & &  \int_k 
\left\{ \sum_{\mu,\nu}
 i\, tr \left( \frac{(\gamma_0 + 1)}{4}
 \frac{d}{dp_0} \left [ V_\mu \;
\frac{\overline{G}_0}{a_t} \;
V_\nu \right]_{p_0=i\mone} \; D_{\mu\nu}\; Z_2^{(0)} \right)
 \; - \; {\cal F}_{sub}(k,m_{\it eff},\Lambda,\lambda)  \right\} \nl
 & & \quad + \quad F(m_{\it eff},\Lambda,\lambda),
\end{eqnarray}
with
\be
\label{addback}
 F(m_{\it eff},\Lambda,\lambda) = 
 \int_k  {\cal F}_{sub}(k,m_{\it eff},\Lambda,\lambda)
\ee
and
\be
\int_k \equiv \intk .
\ee
Explicit forms for ${\cal F}_{sub}(k,m_{\it eff},\Lambda,\lambda)$ and 
$F(m_{\it eff},\Lambda,\lambda)$ are given in the Appendix.  $\lambda$ is 
a gluon mass introduced to regulate IR divengences.  ${\cal F}_{sub}$ 
has been constructed to match the same IR divergence as the first term 
 inside the integral in (\ref{z2sub}). As a result the integral 
becomes independent of $\lambda$.  The other condition on ${\cal F}_{sub}$ is 
that the integral (\ref{addback})  be easy to do analytically. 
 The simplest approach 
is to use a continuum self energy expression for ${\cal F}_{sub}$ with 
an appropriate choice for the mass parameter $m_{\it eff}$.  The need to 
adjust $m_{\it eff}$ to optimize matching of the IR behaviours 
in ${\cal F}_{sub}$ and the lattice integrand,  was emphasized 
in reference \cite{kura} and following that work we find 
\be
\label{meff}
a_t m_{\it eff} = \sinh(a_t\mone) \frac{\cosh(a_t\mone) + \chi \sinh(a_t\mone)}
{1 + \chi \sinh(a_t\mone)} .
\ee
The same ${\cal F}_{sub}$ and $m_{\it eff}$ work for both the clover and D234 
quark actions since the IR structure 
of the two theories agree.
Finally, $\Lambda \leq \pi$ in the above expressions is a cutoff imposed on 
${\cal F}_{sub}$ so that
${\cal F}_{sub} = 0$   for $k^2 > \Lambda^2$. The full expression
 (\ref{z2sub}) must be independent of $\Lambda$.  

\subsection{$\Sigma^{tad}(p)$}
The second contribution to $\Sigma(p)$ is the tadpole contribution 
$\Sigma^{tad}(p)$ coming from the two-gluon emission vertices 
listed in the Appendix.  For quark action $\act^{II}_{D234}$ one 
has
\begin{eqnarray}
a_t \Sigma^{tad}(p) &=& \frac{1}{2} \left[i \gamma_0 \sin(a_t p_0) - 
\chi \,\cos(a_tp_0) \right] \int_k D_{00}    \nl
 & & + \; \frac{1}{2 \chi} \, \frac {1}{3} \sum_j \int_k D_{jj} 
\left\{ i \gamma_j 
\left [ (3 + C_3) \sin(a_sp_j) - 2 C_3 \sin(2 a_s p_j) 
\cos(\frac{k_j}{2}) \right] \right.  \nl
 & & - \left. \left [(3 + C_4) \cos(a_sp_j) - C_4 \cos(2 a_s p_j) 
\cos^2(\frac{k_j}{2}) \right] 
\right \} .
\end{eqnarray}
$\Sigma^{tad}(p)$ in the case of $\act_{clover}$ is obtained by setting 
$C_3 = C_4 = 0$ in the above expression.  The appropriate traces and 
derivatives with respect to external momenta can be carried out
immediately and one has
\begin{eqnarray}
\label{trtad1}
 & & - tr \left\{ \frac{(\gamma_0 + 1)}{4} a_t \Sigma^{tad} \right\}_
{p=(i\mone,\vec{0})}  \nl
 & & = \left\{ \begin{array}{l}
  \frac{1}{2}[\sinh(a_t\mone) + \chi \cosh(a_t\mone) ] \int_k D_{00} 
 + \frac{1}{2 \chi}\sum_j \int_k D_{jj}  \qquad \qquad \qquad \quad 
 \; \act_{clover} \\
                  \\
  \frac{1}{2} [\sinh(a_t\mone) + \chi \cosh(a_t\mone) ] \int_k D_{00} 
 + \frac{1}{6 \chi}\sum_j \int_k D_{jj}[4 - \cos^2(\frac{k_j}{2})] 
 \qquad \act^{II}_{D234}  \\
          \end{array} \right.   \\
 & & \nl
 & & \nl
\label{trtad2}
 & &  i \; tr \left\{ \frac{(\gamma_0 + 1)}{4} \frac{d}{dp_0}
 \Sigma^{tad} \right\}_{p=(i\mone,\vec{0})}  =  
- \frac{1}{2}[\cosh(a_t\mone) + \chi \sinh(a_t\mone) ] \int_k D_{00} \\
& & \hspace{4.8in} \act_{clover} \;\&\;\act^{II}_{D234} \nl
 & & \nl
 & & \nl
\label{trtad3}
 & &B^{tad}_j - B^{tad}_0  = \left\{ \begin{array}{l}
  \frac{1}{2} \int_k D_{jj}  \; - \;
  \frac{ 1}{2} \int_k D_{00} 
 \qquad \qquad \qquad \qquad \qquad 
 \; \act_{clover} \\
                  \\
 \frac{2}{3} \int_k D_{jj} \sin^2(\frac{k_j}{2})   \; - \;
  \frac{ 1}{2} \int_k D_{00} 
\qquad \qquad  
 \qquad \; \; \act^{II}_{D234}  \\
          \end{array} \right.   
\end{eqnarray}
where $B_j^{tad}$ and $B_0^{tad}$ are the contributions from the 
tadpole diagram to (\ref{bj}) and (\ref{b00}) or (\ref{b0}).
All the integrals  are IR finite and very easy to carry out numerically. 
Contributions from $\Sigma^{tad}$ typically dominate 
over those from $\Sigma^{reg}$ but the bulk if not 
all of it is cancelled by $\Sigma^{t.i.}$. 

\subsection{$\Sigma^{t.i.}(p)$}
The lattice covariant derivatives in the quark actions are tadpole-improved. 
They are listed in the Appendix.  In momentum space one has, for 
instance
\be
\nabla_\mu \rightarrow i\,\sin(a_\mu p_\mu)/u_\mu \approx
 i \, \sin(a_\mu p_\mu)
\, [ 1 + \alpha_s u^{(2)}_\mu ] \; + \; O(\alpha_s^2) ,
\ee
where we have perturbatively expanded
\be
u_\mu = 1 - \alpha_s u^{(2)}_\mu  \; + \; O(\alpha_s^2).
\ee
Even in the absence of the regular and tadpole one-loop diagrams there are 
hence $O(\alpha_s)$ terms in the quark propagator. We denote the 
inverse quark propagator with the $u_\mu$'s still in place as 
$\overline{G}^{-1}
_{0,u0}(p)$, so that 
$\overline{G}^{-1}_0(p) \equiv
\overline{G}^{-1}_{0,u0=1}(p)$.  
Through $O(\alpha_s)$ eqn.(\ref{ginv}) can be written as
\be
\overline{G}^{-1} = 
\overline{G}^{-1}_{0,u0}  - \Sigma^{reg} - \Sigma^{tad} 
\equiv
\overline{G}^{-1}_0  - \Sigma^{reg} - \Sigma^{tad} - \Sigma^{t.i.}
\ee
or
\be
\label{sigtimp}
\Sigma^{t.i.} = 
\overline{G}^{-1}_{0,u0=1}  - 
\overline{G}^{-1}_{0,u0}  .
\ee
From the difference in (\ref{sigtimp}) one sees that one link hops 
bring in factors of $1 - 1/u_\mu \approx - \alpha_s u_\mu^{(2)} $ and 
two link hops
factors of $1 - 1/u^2_\mu \approx - 2 \alpha_s u_\mu^{(2)} $ etc.
Using these rules one finds
\begin{eqnarray}
\label{sigu0}
 a_t \, \Sigma^{t.i.}(p)\; &=& \; \alpha_s u^{(2)}_t \, [- i \gamma_0 \sin(a_t p_0) 
 + \chi \cos(a_tp_0)] \; + \;  \nl
 & & \alpha_s u^{(2)}_s \, \frac{1}{3 \chi}
\sum_j\left\{ -i \gamma_j \, [(3 + C_3) \sin(a_s p_j)  - C_3\sin(2a_sp_j) ]
\right. \nl
& & \qquad  +  \qquad \qquad \left. [(3 + C_4) \cos(a_sp_j)
 - \frac{C_4}{2} \cos(2a_sp_j)] \right\} .
\end{eqnarray}
The relevant traces and derivatives become
\begin{eqnarray}
\label{tru01}
 & & - tr \left\{ \frac{(\gamma_0 + 1)}{4} a_t \Sigma^{t.i.} \right\}_
{p=(i\mone,\vec{0})}  \nl
 & & = \left\{ \begin{array}{l}
- [\sinh(a_t\mone) + \chi \cosh(a_t\mone) ] \; \alpha_s u^{(2)}_t
- \frac{3}{ \chi} \; \alpha_s u^{(2)}_s  \qquad \qquad \qquad \quad 
 \; \act_{clover} \\
                  \\
- [\sinh(a_t\mone) + \chi \cosh(a_t\mone) ] \; \alpha_s u^{(2)}_t
-  \frac{1}{\chi}\, \frac{7}{2} \, \alpha_s u^{(2)}_s
\; \; \qquad \qquad  \qquad \act^{II}_{D234}  \\
          \end{array} \right.   \\
 & & \nl
 & & \nl
\label{tru02}
 & &  i \; tr \left\{ \frac{(\gamma_0 + 1)}{4} \frac{d}{dp_0}
 \Sigma^{t.i.} \right\}_{p=(i\mone,\vec{0})}  =  
 [\cosh(a_t\mone) + \chi \sinh(a_t\mone) ] \; \alpha_s u^{(2)}_t \\
 & & \hspace{4.5in} \act_{clover} \;\&\;\act^{II}_{D234} \nl
 & & \nl
 & & \nl
\label{tru03}
 & &B^{t.i.}_j - B^{t.i.}_0  = \left\{ \begin{array}{l}
       \alpha_s  \, ( u^{(2)}_t - u^{(2)}_s)
 \qquad \qquad \qquad \qquad \qquad 
 \; \act_{clover} \\
                  \\
        \alpha_s \, ( u^{(2)}_t - \frac{2}{3} u^{(2)}_s )
\qquad \qquad  
\; \; \; \qquad \qquad \; \; \act^{II}_{D234}  \\
          \end{array} \right.   
\end{eqnarray}
The Landau mean link definition of $u_\mu$ is given by
\be
u_\mu \equiv \langle \frac{1}{3} Tr U_\mu \rangle_{\alpha_g = 0}  \approx 1 - 
 \alpha_s u^{(2)}_\mu = 1 - \frac{1}{2} \int_k D_{\mu\mu}(\alpha_g = 0) .
\ee
If one evaluates $\Sigma^{tad}$ in Landau gauge then 
(\ref{trtad1}) \& (\ref{tru01}) ,
(\ref{trtad2}) \& (\ref{tru02}) and
(\ref{trtad3}) \& (\ref{tru03}) cancel for $\act_{clover}$. 
(\ref{trtad2}) \& (\ref{tru02}) also cancel for $\act^{II}_{D234}$ and 
for the other two traces cancellation is almost complete. The 
difference between contributions from $\Sigma^{tad}$ and $\Sigma^{t.i.}$ 
would go away if one replaces
 $\cos^2(k/2)$
and  $\sin^2(k/2)$ by their averages $1/2$.  Hence, it is 
easy to see in this calculation how tadpole improving terms in the lattice 
action eliminates lattice artifact contributions in perturbation theory.

\section{Results}
\label{sec:results}

\noindent
In this section we summarize results for the one-loop coefficients, 
$a_s M_{1,sub}^{(1)}$, $Z_2^{(1)}$ and $C_0^{(1)}$ for mass, wave function 
and speed of light renormalizations respectively.  These follow from 
equations (\ref{m11b}), (\ref{z2inv1}) and (\ref{c0}) - (\ref{b0}) and each 
has, as explained in the previous section, contributions from 
regular, tadpole and t.i. diagrams.  The numbers in our Tables are 
coefficients of $\alpha_s$. The Landau mean link definition of $u_\mu$ is 
used throughout to implement tadpole improvement.

\subsection{  $a_sM_{1,sub}^{(1)}$ }

\noindent
In Table II we present results for $a_s M_{1,sub}^{(1)}$ for action 
$\act^A$ for several values of $a_s M_1^{(0)}$.  We list separately 
contributions from $\Sigma^{reg}$, $\Sigma^{tad}$ and $\Sigma^{t.i.}$.  
The fourth column gives the gauge invariant combination (reg + tad) and 
the sixth column gives
$a_s M_{1,nosub}^{(1)} \equiv$ (reg + tad + t.i.), 
the full tadpole improved one-loop correction before subtraction. 
Carrying out the subtraction according to (\ref{m11b}), one obtains 
$a_s \monesub$ which is given in the last column 
\be
\label{m1sub}
a_s \monesub = a_s M_{1,nosub}^{(1)} - \frac{a_sM_{1,nosub}^{(1)}(0)}
{\chi \sinh(a_t\mone) + \cosh(a_t \mone)} .
\ee
All our calculations have been carried out for two values of the gauge fixing 
parameter $\alpha_g$, 1.0 and 0.0.  Table II lists both sets of results and 
one sees that gauge invariant quantities are independent of $\alpha_g$ 
within numerical integration errors ( which we take to be at the 
$\pm 0.003$ to $\pm 0.006$ level depending on the mass). 
 Our results for $a_s\mone = 0$ agree with those from 
\cite{aoki}.  

\vspace{.1in}
\noindent
We plot $a_s \monesub$ versus $a_s \mone$ in Fig. 1 .  One sees that 
the mass dependence is smooth and that one reaches saturation rapidly 
already around $a_s \mone \sim 3.0 - 5.0$.  We also compare with
non-tadpole improved results for which 
 the curve saturates around 1.827 rather 
than around 1.077.  In considering the large mass limit it is useful to 
note that the factor $Z_2^{(0)} = 1/[\chi \sinh(a_t \mone) +
 \cosh(a_t \mone)]$ appearing in (\ref{m11b}) and (\ref{z2inv1}) vanishes 
exponentially in this limit.  The only terms that survive into the static 
limit are those where $Z_2^{(0)}$ is multiplied by an exponentially 
increasing function of $a_s \mone$.  It is easy to see, for instance, 
 that the subtraction 
term in (\ref{m1sub}) or the spatial tadpoles in (\ref{trtad1}) become 
irrelevant in the static limit.  Furthermore $a_s \monesub$ becomes
identical for $\act_{Wilson}$, $\act_{clover}$ and $\act^{II}_{D234}$ 
in this limit and the only difference between the current calculations 
and those of reference \cite{eichhill} resides in the glue action (we have 
verified that by switching to the unimproved Wilson glue action 
results of \cite{eichhill} are reproduced).

\vspace{.1in}
\noindent
In Tables III, IV and V we summarize results for $a_s\monesub$ for 
the other actions listed in Table I.  We also list the combination (reg + tad)
for each case.  If one chooses to implement tadpole improvement differently 
from what we have done here or decides not to tadpole improve, then 
$a_s \monesub$ can be calculated straightforwardly from (reg + tad) and 
the formulas presented in this paper.  For action $\act^C$ we list only 
massless results for reasons explained in section II.  The 
dispersion relations of this and similar actions  are discussed in reference 
\cite{klassen}.   Our choices for anisotropy values in Tables IV and V, 
were dictated in part with an eye towards practical numerical 
simulations.  Values such as $\chi=3.6$ and $\chi=5.3$ were taken from 
recent work on nonperturbative determinations of the renormalized 
anisotropy in 
pure glue theory \cite{ron}. 
For one value $a_s \mone = 1 $ we plot $a_s \monesub$ versus
 $\chi$ in Fig. 2 using the action $\act^D$.  From Tables IV and V and 
from Fig. 2 one sees that the dependence of $a_sM_{1,sub}^{(1)}$ on $\chi$ 
is very mild.  Effects of tadpole improvement are significant 
 only for small values of $\chi$.  This is because due to cancellations 
in the subtraction of equation (\ref{m1sub}) only the temporal tadpole
and the temporal Landau link term $u_t^{(2)}$ contribute to 
$a_sM^{(1)}_{1,sub}$ and both these become small as $\chi$ increases.

\vspace{.1in}
\noindent
Finally we mention that the one-loop expression for the critical 
bare mass $m_c$ is given by
\be
a_tm_c = - \frac{\alpha_s}{\chi} a_s M^{(1)}_{1,nosub}(0) \; + \;
 O(\alpha_s^2).
\ee

\subsection{  $Z_2^{(1)}$ }

\noindent
Starting with (\ref{z2inv1}) we define 
\be
Z_2 = Z_2^{(0)} \, [ \, 1 \, + \, \alpha_s \,(Z_2^{(1)} + 
Z_2^{(1)IR}) \; + \; O(\alpha_s^2)]
\ee
with
\be
Z_2^{(1)IR} = 
    \left\{ \begin{array}{l}
             \frac{1}{3 \pi} \,  
[\;\;  1 \;+\; (\alpha_g -1 )\, ] \; \ln(\lambda^2) \qquad \qquad m = 0 \\
\\
             \frac{1}{3 \pi} \,  
[ - 2 \;+\; (\alpha_g -1 )\, ] \; \ln(\lambda^2) 
  \qquad \qquad m > 0 \\
                                                  \end{array} \right.   
\ee
$\lambda$ is the gluon mass in units of $1/a_s$.  
It is the coefficient of $\alpha_s$ after factoring out 
$Z_2^{(0)}$ that has the same IR  $\ln(\lambda)$ and $\ln(am)$ 
structure as the continuum wave function renormalization constant.  
From (\ref{z2inv1}) one also sees that there are two contributions to 
$Z_2^{(1)}$, one coming from the $d/dp_0$ derivative term 
and the second from the expansion of $M_1$.  Accordingly we write
\be
\label{z2one}
Z_2^{(1)} = Z_{2,dp_0}^{(1)} + Z_{2,M_1}^{(1)} .
\ee
In the literature $Z_{2,M_1}^{(1)}$ is not always included as part 
of the definition of $Z_2^{(1)}$. $Z_{2,dp_0}^{(1)}$ alone with 
unimproved Wilson glue goes over in the large mass limit
 to the wave function 
renormalization of reference \cite{eichhill}.
Including $Z_{2,M_1}^{(1)}$ leads to the static result of reference 
\cite{boupene} which has been used in many subsequent static calculations, 
 for instance in \cite{borrpit}.  
This latter static value also corresponds to the large mass limit of 
the one-loop $Z_2$ calculated in many versions of NRQCD actions \cite{colin}.

\vspace{.1in}
\noindent
Table VI presents results  for $Z_{2,dp_0}^{(1)}$ 
and the full $Z_2^{(1)}$ for the action $\act^A$.  Again we agree 
with reference \cite{aoki} for $ m = 0$. However, 
one notices that the massive data do not tend towards the massless 
result as $a_s\mone$ decreases.  This is because our massive numbers include 
$\ln(am)$ contributions which will eventually diverge, 
whereas in the massless theory we have set the fermion mass identical 
to zero from the beginning.
This leads to different IR structure for the two theories (see Appendix 
B for some further discussions).  In a matching calculation 
one will be looking at differences between the lattice and continuum $Z_2$. 
As long as IR divergences are handled in the same manner in 
the lattice and continuum evaluations, one should not run into any 
problems and the $m \rightarrow 0$ limit
 should be smooth.  For instance, using 
dimensional regularization in the $\overline{MS}$ scheme one finds 
in Feynman gauge the UV finite continuum results
\be
Z_2^{(1)\,cont.} = 
    \left\{ \begin{array}{l}
             \frac{1}{3 \pi} \, \left[ \,
             \ln(\frac{\hat{\lambda}^2}{\mu^2}) + \frac{1}{2}\, \right] 
             \;\; \qquad \qquad \qquad \qquad  m = 0 \\
                                                                 \\
             \frac{1}{3 \pi} \, \left[ \,
             \ln(\frac{m^2}{\mu^2}) + 2 \ln(\frac{m^2}
             {\hat{\lambda}^2}) - 4 \, \right] 
             \qquad \qquad m > 0 \\
                                                  \end{array} \right.   
\ee
Taking the difference between continuum and lattice wave function 
renormalization constants, it makes sense to consider the 
following subtracted $Z_2$ factors.
\be
\label{z2diff}
Z^{(1)}_{2,diff} = Z_{2,dp_0}^{(1)} 
                  -  \left\{ \begin{array}{l}
             \frac{1}{3 \pi} \, \frac{1}{2}  
             \;\; \; \;\qquad \qquad \qquad \qquad \qquad m = 0 \\
                                                                 \\
             \frac{1}{3 \pi} \, [ \,
             3 \ln(a_s\mone)^2- 4  \,]
             \qquad \qquad m > 0 \\
                                                  \end{array} \right.   
\ee
Numbers for $Z_{2,diff}^{(1)}$ are given in Table VII and 
one sees that the 
$m \rightarrow 0$ behaviour is smooth.  

\vspace{.1in}
\noindent
In Tables VIII through XII we present $Z_{2,dp_0}^{(1)}$ and $Z_2^{(1)}$ 
for other actions.  One does not find any dramatic changes with differing 
actions and/or anisotropy.  The IR subtractions of Appendix B worked well 
in all actions for $a_s\mone < \; \sim 5$.  For larger masses VEGAS errors 
became large especially for  $\chi = 1$.  Hence, we only present results up to 
$a_s\mone = 5$.  
For $\chi >1 $  problems were less severe in general.
In Tables VI - XII the numerical integration errors are at the 
$\pm 0.02$ level for $a_s\mone = 5.0$ and $\chi <3$, at the $\pm0.006$ level
for $a_s\mone$ = 0.01, 0.05 and 0.10 
and at the $\pm 0.004$ level for all other cases.  A more sophisticated method 
for handling IR divergent integrals appears necessary if accurate 
results are required
 for larger masses.  Many quantities, however,
are close to saturation by the time one reaches $a_s\mone = 5$.  

\subsection{  $C_0^{(1)}$ }

\noindent
For the speed of light renormalization we present the most detailed 
results for action $\act^B$ rather than for $\act^A$ since the former, 
for $\chi > 1$, is genuinely anisotropic.  In Table XIII we list separately 
$regular$ and $tadpole$ diagram contributions, their gauge invariant 
sum $(reg \;+ \;tad) \equiv (C_0^{(1)} \;\; no \;\; t.i.)$ 
and the fully tadpole improved result $(C_0^{(1)} \;\; with \;\; t.i.)$, all 
for action $\act^B$ and at fixed anisotropy $\chi = 4.0$.   $C_0^{(1)}$ 
is independent of $\alpha_g$ within numerical integration errors which are 
the most severe when using (\ref{b00}) for nonzero but small masses. 
 In Figure 3. we plot $C_0^{(1)}$ both 
with and without tadpole improvement versus $a_s\mone$.  Table XIV 
summarizes results for several $\chi$ values with action $\act^B$ and in 
Figure 4. we plot $C_0^{(1)}$ versus $\chi$ for 
fixed $a_s\mone = 1.0$.  One sees that tadpole improvement has significant
 effect and  causes $C_0^{(1)}$ to switch sign for $\chi >1$. 
Among other things this allows for a smooth $\chi \rightarrow 1$ limit.

\vspace{.1in}
\noindent
In Table XV we present results for actions $\act^A$ and $\act^{A^\prime}$.
 In these isotropic actions nontrivial $C_0$ comes about 
because our mass-shell condition $p = (iM_1, \vec{0})$ distinguishes 
between spatial and temporal directions once $M_1 >0$. Table XVI 
summarizes results for action $\act^D$. 
Here tadpole improvement 
does not decrease the magnitude of the one-loop correction, however,
for a wide range of mass values it is still true 
that $C_0^{(1)}$ switches sign for $\chi >1$ and that the $\chi \rightarrow 1$ 
limit becomes smoother after tadpole improvement.

\section{Summary}
\label{sec:summary}
We have carried out one-loop perturbative renormalization of the 
fermion rest mass $M_1$, wave function renormalization $Z_2$ and 
the speed of light $C_0$ for a range of highly improved 
actions on isotropic and anisotropic lattices.  We find that the 
dependence of the one-loop coefficients on the anisotropy parameter 
$\chi = a_s/a_t$ and on the tree-level mass parameter $a_s \mone$ 
is mild,  especially after tadpole improvement of the actions.  
Furthermore, none of the coefficients are particularly large.  
 $M_1$ and $C_0$ exhibit smooth behaviour as one approaches
 the massless, large mass, $\chi \rightarrow 1$ 
and large $\chi$ limits.  This also holds for $Z_2$ if more physical 
combinations such as the difference between continuum and lattice 
wave function renormalizations are considered.  The next stage in our 
program would be to extend the present calculations to vertex corrections 
and to matchings
 between continuum and lattice currents and other multi-fermion operators.

\section{Acknowledgments}
This research is supported by grants from the US Department of Energy,
 DE-FG02-91ER40690, and from the Graduiertenkolleg.
  The authors thank Peter Lepage for many 
useful conversations.
S.G. gratefully acknowledges a grant given by the Max Kade Foundation. 
  J.S. thanks Cornell University for its hospitality 
while the project was being initiated and the theoretical physics group at 
the University of Glasgow where part of the work was carried out.
Support from a UK PPARC Visiting Fellowship PPA/V/S/1997/00666
 is gratefully acknowledged.

\appendix

\section{Definitions and Feynman rules}
\label{sec:feynrules}
In this Appendix we summarize definitions for various terms in the 
lattice actions and present Feynman rules for gauge and quark propagators 
and for vertex functions.

\vspace{.2in}
\noindent
\underline{ Covariant Derivatives Acting on Quark Fields }

\begin{eqnarray}
\nabla_\mu\, \Psi(x) &=& \frac{1}{2} \, \frac{1}{u_\mu} \,
 [ U_\mu(x) \Psi(x+a_{\mu})
\; - \; U^\dagger_\mu(x - a_{\mu}) \Psi(x - a_{\mu}) ]  \\
\nabla^{(2)}_\mu\, \Psi(x) &=&  \frac{1}{u_\mu} \,
 [ U_\mu(x) \Psi(x+a_{\mu})
\; + \; U^\dagger_\mu(x - a_{\mu}) \Psi(x - a_{\mu}) ] \; - \; 2\,\Psi(x) \\
\nabla^{(3)}_\mu\, \Psi(x) &=& 
\frac{1}{2} \, \frac{1}{u^2_\mu} \, [ U_\mu(x) U_\mu(x +  a_{\mu})
 \Psi(x+ 2a_{\mu})
\; - \; U^\dagger_\mu(x - a_{\mu}) U^\dagger_\mu(x - 2 a_{\mu})
 \Psi(x - 2a_{\mu}) ]  \nl
 & & - \, \frac{1}{u_\mu} \, [ U_\mu(x) \Psi(x+a_{\mu})
\; - \; U^\dagger_\mu(x - a_{\mu}) \, \Psi(x - a_{\mu}) ]  \\
\nabla^{(4)}_\mu\, \Psi(x) &=& 
 \frac{1}{u^2_\mu} \, [ U_\mu(x) U_\mu(x +  a_{\mu})
 \Psi(x+ 2a_{\mu})
\; + \; U^\dagger_\mu(x - a_{\mu}) U^\dagger_\mu(x - 2 a_{\mu})
 \Psi(x - 2a_{\mu}) ]  \nl
 & & -\,4 \, \frac{1}{u_\mu} \, [ U_\mu(x) \Psi(x+a_{\mu})
\; + \; U^\dagger_\mu(x - a_{\mu}) \, \Psi(x - a_{\mu}) ] 
\; + \; 6\,\Psi(x) 
\end{eqnarray}

\vspace{.2in}
\noindent
\underline{ Field Strength Tensors }

\vspace{.1in}
\noindent
For the unimproved $F_{\mu\nu}$ of the clover action we use
\begin{eqnarray}
  F_{\mu\nu}(x) &=& {1\over 2i}
\left( \Omega_{\mu\nu}(x)-\Omega^\dagger_{\mu\nu}(x)\right), \nonumber\\
\Omega_{\mu\nu}(x) &=&  {1\over 4u_\mu^2 u_\nu^2}\ \sum_{{\lbrace(\alpha,\beta)
   \rbrace}_{\mu\nu}}\!\!U_\alpha(x)U_\beta(x\!+\!a_\alpha)
   U_{-\alpha}(x\!+\!a_\alpha\!+\!a_\beta)U_{-\beta}(x\!+\!a_\beta),
\label{fieldstrength}
\end{eqnarray}
with $\lbrace(\alpha,\beta)\rbrace_{\mu\nu} = \lbrace (\mu,\nu),
(\nu,-\mu), (-\mu,-\nu),(-\nu,\mu)\rbrace$ for $\mu\neq\nu$ and 
$U_{-\mu}(x + a_{\mu}) \equiv U^\dagger_\mu(x)$.  
The $O(a^2)$ improved field strength tensor of the D234 actions is 
\begin{eqnarray}
& &\tilde{F}_{\mu\nu}(x) = \frac{5}{3}F_{\mu\nu}(x) \nl
& & \quad \; - \; \frac{1}{6} \left[\,\frac{1}{u_\mu^2}
(\,U_\mu(x)F_{\mu\nu}(x+a_\mu)U^\dagger_\mu(x) + U^\dagger_\mu(x-a_\mu)
F_{\mu\nu}(x-a_\mu)U_\mu(x-a_\mu)\,)  
\; - \;  (\mu \leftrightarrow \nu)\,\right]  \nl
& & \quad \; + \; \frac{1}{6} \, ( \frac{1}{u_\mu^2} + \frac{1}{u_\nu^2} 
- 2 ) \, F_{\mu\nu}(x).
\end{eqnarray}
The last term ensures that factors of $1/u_\mu$ are correctly removed 
from those contributions to $U  F_{\mu\nu}  U^\dagger$ and $U^\dagger 
 F_{\mu\nu} U$ that end up being 
four link objects rather than six link ones. 
  In a one-loop calculation, however, one can set $u_\mu = 1$ 
everywhere in the definition of the field strength tensor and this correction 
term is irrelevant.

\vspace{.1in}
\noindent
Both the above covariant derivatives and the field strength tensor are 
dimensionless.  Factors of $1/a_t$ and $1/a_s$
are inserted explicitly where necessary such as in (\ref{sd234a}).

\vspace{.2in}
\noindent
\underline{ Gauge Propagator }

\vspace{.1in}
\noindent
The isotropic Symanzik improved gauge action has been discussed 
quite extensively in the literature \cite{weisz}.  Here we summarize 
formulas for the anisotropic generalization.
We start from the gauge actions $\act^I_G$ or $\act^{II}_G$ and add to 
it a gauge fixing term
\begin{eqnarray}
\label{sgfa}
\act_{gf} &=& \frac{1}{2 \alpha_g} \, a_s^3 a_t \sum_x \left[ \frac{1}{a_t}
\, \partial_t A_t + \frac{1}{a_s} \,\sum_j \partial_j A_j \right]^2 \\
\label{sgfb}
 &=& \frac{1}{2 \alpha_g} \, \frac{1}{\chi} \, \sum_x \left[
\chi^2 \partial_t(a_t A_t) + \sum_j \partial_j(a_sA_j) \right]^2 ,
\end{eqnarray}
with $\partial_\mu A_\mu(x) \equiv A_\mu(x + a_\mu/2) - 
A_\mu(x - a_\mu/2)$.  Equation (\ref{sgfb}) expresses $\act_{gf}$ 
in terms of dimensionless gauge fields $a_\mu A_\mu$.  It is 
convenient to do so, especially since $\act^{I,II}_G$ are already in 
dimensionless form with factors of $\chi$ and $1/\chi$ properly 
put in place.  If $\bar{A}_\mu(k)$ is the Fourier transform of 
$(a_\mu A_\mu)$, the quadratic terms in the gauge action become
\be
\act^{(0)I,II}_G + \act_{gf} = \frac{1}{2} \, \sum_{\mu\nu} 
\int_{-\pi}^\pi \frac{d^4 k}{(2 \pi)^4} \left( \bar{A}_\mu(k) 
 \,M_{\mu\nu}(k) \, \bar{A}_\nu(-k) \right) ,
\ee
where
\begin{eqnarray}
M_{00} &=& \chi \, \left[ \frac{\chi^2}{\alpha_g} \hkt^2 \; + \; \sum_j
\hkj^2 \, q_{0j} \right]   \\
M_{jj} &=& \frac{1}{\chi} \, \left[ \frac{1}{\alpha_g} \hkj^2 \; + \; 
\chi^2 \, \hkt^2 \, q_{0j} \; + \; \sum_{l\neq j}\hat{k}_l^2 \, q_{lj}
 \right]   \\
M_{i \neq j} &=& \frac{1}{\chi} \, \left[ \frac{1}{\alpha_g} 
 \hat{k}_i \hkj \; - \;  \hat{k}_i \hkj \, q_{ij} \right]   \\
M_{0 j} = M_{j 0}  &=& \chi \, \left[ \frac{1}{\alpha_g} 
 \hkt \hkj \; - \;  \hkt \hkj \, q_{0j} \right]  
\end{eqnarray}
and
\be
\hat{k}_\mu \equiv 2\,\sin(\frac{k_\mu}{2}) .
\ee
The $q_{\mu\nu}$ need to be specified only for $\mu \neq \nu$ and one has
\begin{eqnarray}
q_{\mu\nu} &=& 1 \; - \; c^G_1 \, (\hat{k}^2_\mu + \hat{k}^2_\nu) 
\qquad \mu \neq \nu \qquad \qquad \act^I_G \\
 & & \nl
q_{ij} &=& 1 \; + \; \frac{1}{12} \, (\hat{k}^2_i + \hat{k}^2_j) 
\qquad i \neq j 
 \hspace{.65in} \act^{II}_G \nl
q_{0j} &=& 1 \; + \; \frac{1}{12} \, \hat{k}^2_j   
 \hspace{1.8in} \act^{II}_G 
\end{eqnarray}
We have inverted the $4 \times 4$ matrix $M_{\mu\nu}$ using Mathematica 
keeping $q_{\mu\nu}$ general.  
For both gauge actions, $\act^I_G$ and $\act^{II}_G$ the free gauge 
propagator has the structure
\be
\label{dmunu}
D_{\mu\nu}(k) = M^{-1}_{\mu\nu}
 = \frac{1}{(\hat{k}^2)^2} \left[ \alpha_g \hat{k}_\mu
\hat{k}_\nu \chi + \frac{f_N^{\mu\nu}(\hat{k}_\rho,q_{\rho\sigma},\chi)}
{f_D(\hat{k}_\rho,q_{\rho\sigma},\chi)} \right] ,
\ee
with
\be
\hat{k}^2 = \chi^2 \hkt^2 + \sum_j \hkj^2 .
\ee
The term proportional to the gauge fixing parameter $\alpha_g$ has the 
familiar form
\be
\alpha_g \,\frac{\hat{k}_\mu \hat{k}_\nu}{(\hat{k}^2)^2} \, \chi = 
\alpha_g \,\frac{a_\mu a_\nu}{a_s^3a_t} \, \frac{\hat{k}_\mu \hat{k}_\nu/
(a_\mu a_\nu)} {[(\hkt/a_t)^2 + \sum_j(\hkj/a_s)^2]^2} 
\ee
with the conversion factor $a_\mu a_\nu/a_s^3a_t$ mentioned in (\ref{dtilde}).
This factor results because we are looking at the propagator for 
dimensionless gauge fields $a_\mu A_\mu$ and because we carried out 
a dimensionless Fourier transform.
The second term in (\ref{dmunu}) is much more complicated.  If 
one writes
\begin{eqnarray}
f_N^{00}(\hat{k}_\rho,q_{\rho\sigma},\chi) &=& 
\frac{1}{\chi}\,\tilde{f}_N^{00}  \\
f_N^{jj}(\hat{k}_\rho,q_{\rho\sigma},\chi) &=& 
\chi\,\tilde{f}_N^{jj}  \\
f_N^{i \neq j}(\hat{k}_\rho,q_{\rho\sigma},\chi) &=& 
\chi\,\hat{k}_i \hkj
\,\tilde{f}_N^{i \neq j}  \\
f_N^{0j}(\hat{k}_\rho,q_{\rho\sigma},\chi) &=& 
\chi\,\hkt \hkj
\,\tilde{f}_N^{0j}  
\end{eqnarray}
one can show that $f_D$ and all the $\tilde{f}_N^{\mu\nu}$ are 
 functions only of $(\chi \hkt)^2,\hkj^2, q_{\rho\sigma}$ with no 
other $\chi$ dependence or odd powers of $\hat{k}_\rho$. 
We have not shown color indices in the above expressions.  The gluon 
propagator is diagonal in color.

\vspace{.2in}
\noindent
\underline{ Quark Propagator }

\vspace{.1in}
\noindent
The inverse free quark propagator for $\act^I_{D234}$ is given by
\be
a_t\, \overline{G}^{-1}_0(k) = i \gamma_0 K_0(k_0) \sin(k_0) 
\, + \, i \frac{C_0}{\chi} \,\sum_j \gamma_j K_j(k_j) \sin(k_j)
\, + \, \Omega(k_0,\vec{k})
\ee
with
\begin{eqnarray}
K_0 &=& 1 \,+\, \frac{C_{3t}}{3}  \, - \, \frac{C_{3t}}{3} \cos(k_0)  \\
K_j &=& 1 \,+\, \frac{C_{3}}{3}  \, - \, \frac{C_{3}}{3} \cos(k_j)
\end{eqnarray}
and
\begin{eqnarray}
\Omega &=& \, \chi \left[ 2 (1 + \frac{C_{4t}}{3}) 
\sin^2(\frac{k_0}{2}) \, - \, \frac{C_{4t}}{6} \sin^2(k_0) \right] \nl
 && \, \frac{1}{\chi}\sum_j \left[ 2 (1 + \frac{C_{4}}{3}) 
\sin^2(\frac{k_j}{2}) \, - \, \frac{C_{4}}{6} \sin^2(k_j) \right] 
\; + \; a_t \, m .
\end{eqnarray}
Propagators for the other quark actions can be obtained by setting the 
appropriate $C_{i(t)}$ equal to zero.  Quark propagators are diagonal in 
color.

\vspace{.2in}
\noindent
\underline{ Vertex Functions }

\vspace{.1in}
\noindent
In deriving the one- and two-gluon emission vertices we have used the 
method described in \cite{colin}.  We list again results only for 
$\act^I_{D234}$.  Those for other quark actions follow trivially.
The general form for a single gluon emission vertex is
\be
V_\mu(k^\prime,k) \equiv -i \gamma_\mu W_\mu \,-\, W'_\mu \,-\, 
\sum_\nu \sigma_{\nu\mu} \, W''_{\nu\mu}
\ee
where $\mu$ is the polarization of the emitted gluon, $k^\prime$ the 
momentum of the outgoing quark and $k$ the momentum of the incoming 
quark.  We suppress the color factor $T^a_{bc}$ which should 
multiply each of the above terms.  Using the variables
\be
k^\pm_\mu \equiv \frac{1}{2} ( k^\prime \pm k)_\mu
\ee
one has
\begin{eqnarray}
W_0 &=& \;\;\;(1 + \frac{C_{3t}}{3}) \cos(k^+_0) \,-\, \frac{C_{3t}}{3} 
\cos(2k^+_0) \, \cos(k^-_0)  \\
W_j &=& \frac{C_0}{\chi} \left[(1 + \frac{C_{3}}{3}) \cos(k^+_j) \,-\,
 \frac{C_{3}}{3} \cos(2k^+_j) \, \cos(k^-_j) \right] \\
W'_0 &=& \chi \left[(1 + \frac{C_{4t}}{3}) \sin(k^+_0) \,-\,
 \frac{C_{4t}}{6} \sin(2k^+_0) \, \cos(k^-_0) \right] \\
W'_j &=& \frac{1}{\chi} \left[(1 + \frac{C_{4}}{3}) \sin(k^+_j) \,-\,
 \frac{C_{4}}{6} \sin(2k^+_j) \, \cos(k^-_j) \right] 
\end{eqnarray}
and
\begin{eqnarray}
W''_{j0} &=& \frac{1}{2} \sin(2k^-_j) \cos(k^-_0) \, \frac{1}{3} \,
[5 - \cos(2k^-_j) - \cos(2k^-_0) ]  \\
W''_{0j} &=& \frac{1}{2} \sin(2k^-_0) \cos(k^-_j) \,\frac{1}{3} \,
[5 - \cos(2k^-_j) - \cos(2k^-_0) ]  \\
W''_{ij} &=& \frac{1}{2}\frac{1}{\chi} \sin(2k^-_i) \cos(k^-_j) \,\frac{1}{3}\,
[5 - \cos(2k^-_i) - \cos(2k^-_j) ]  .
\end{eqnarray}
For the clover action the factor $\frac{1}{3}[5 - \cos(2k^-_\mu) - 
\cos(2k^-_\nu)]$ in $W''_{\mu\nu}$ should be replaced by $1$.

\vspace{.1in}
\noindent
For the two-gluon emission vertex we do not present the most general 
result, but restrict ourselves to those terms necessary for 
the tadpole diagram $\Sigma^{tad}$. For instance, the $\sigma_{\mu\nu} 
F_{\mu\nu}$ term does not contribute to the tadpole diagram.  We also 
omit terms that vanish upon symmetrizing between the two gluons. If 
$V^{(2)}_{\mu_1 \mu_2}(k^\prime,k,q_1,q_2)$ stands for the emission vertex for 
gluons of momentum $q_i$ and polarization $\mu_i$, with 
$k_\mu = k^\prime_\mu + q_{1,\mu} + q_{2,\mu}$, one has
\begin{eqnarray}
V^{(2)}_{00} &=& \frac{i}{2} \gamma_0 \left[ (1 + \frac{C_{3t}}{3})
\sin(k^+_0) \,-\, \frac{2}{3} C_{3t} \sin(2k^+_0) 
\cos(\frac{q_{1,0}}{2}) \cos(\frac{q_{2,0}}{2}) \right]  \nl
 & & - \frac{\chi}{2} \left[ (1 + \frac{C_{4t}}{3})
\cos(k^+_0) \,-\, \frac{1}{3} C_{4t} \cos(2k^+_0) 
\cos(\frac{q_{1,0}}{2}) \cos(\frac{q_{2,0}}{2}) \right]  \\
& &  \nl
V^{(2)}_{jj} &=& \frac{i}{2} \frac{C_0}{\chi} \gamma_j
 \left[ (1 + \frac{C_{3}}{3})
\sin(k^+_j) \,-\, \frac{2}{3} C_{3} \sin(2k^+_j) 
\cos(\frac{q_{1,j}}{2}) \cos(\frac{q_{2,j}}{2}) \right]  \nl
 & & - \frac{1}{2 \chi}  \left[ (1 + \frac{C_{4}}{3})
\cos(k^+_j) \,-\, \frac{1}{3} C_{4} \cos(2k^+_j) 
\cos(\frac{q_{1,j}}{2}) \cos(\frac{q_{2,j}}{2}) \right]  .
\end{eqnarray}
The color factor for these vertex functions is $(T^{a_1} T^{a_2})_{bc}$.

\section{IR Subtractions}
\label{sec:irsub}
In this Appendix we list the IR subtraction, ${\cal F}_{sub}$ of 
equation (\ref{z2sub}), necessary to control numerical integration 
of IR divergent integrals.  A gluon mass, $\lambda/a_s$,
 is introduced into $D_{\mu\nu}$ 
by replacing the first factor in (\ref{dmunu}) by 
\be \frac{1}{(\hat{k}^2)^2} \; \rightarrow \; \frac{1}{\hat{k}^2} \,
 \frac{1}{\hat{k}^2 + \lambda^2} .
\ee
The lattice wave function renormalization $Z_2$ must reproduce 
the same IR divergence structure as in continuum QCD.  For $Z_2^{-1}$ at 
one-loop the IR divergence is
\begin{eqnarray}
\label{contir}
& &\frac{\alpha_s}{3 \pi} \,
[ - 1 \;-\; (\alpha_g -1 )\, ] \; \ln(\lambda^2) \qquad \qquad m = 0 \nl
& &\frac{\alpha_s}{3 \pi} \,
 [\;\; \;2 \;-\; (\alpha_g -1 )\, ] \; \ln(\lambda^2) \qquad \qquad m > 0 
\end{eqnarray}
We note that by the $m=0$ theory we mean one in which the quark mass has been 
set to zero before taking the limit $ \lambda \rightarrow 0$.  This 
is the usual practice in much of the 
literature on massless lattice perturbation theory.  Alternatively one could 
take the limit $am \rightarrow 0$ and $\lambda \rightarrow 0$ keeping 
$am \geq \lambda$.
 Since we want to compare with some of the massless 
literature with improved glue actions (e.g. \cite{aoki}) we adopt the 
first definition in this article.  In our massive calculations we do not go 
to extremely small masses and have not attempted to isolate $\ln(am)$ 
contributions.

\vspace{.1in}
\noindent
For our VEGAS integrations it was convenient to separate the $d/dp_0$
derivative in (\ref{z2sub}) into two parts
\be
\label{dparts}
 \frac{d}{dp_0} \left[V_\mu \, \frac{\overline{G}_0}{a_t} \, V_\nu 
\right] \equiv
 \frac{d}{dp_0} \left[\frac{VGV_{num}}{VGV_{den}} \right] =
\frac{VGV_{num}^{\prime}}{VGV_{den}} - \frac{VGV_{num}}{(VGV_{den})^2} 
\,VGV_{den}^{\prime} .
\ee
Corresponding to the two parts with derivatives acting on the numerator 
or denominator, respectively, we introduce two separate subtraction 
terms ${\cal F}_{sub}^{num}$ and  ${\cal F}_{sub}^{den}$.  These are 
obtained by calculating the  
self energy diagram in continuum Euclidean perturbation theory 
with an appropriate mass $m_{\it eff}$ and the mass-shell condition 
$p = (i\, m_{\it eff}, \vec{0})$.  
The effective mass follows from expanding the lattice integrand 
in (\ref{z2sub}) about small $k$ and comparing with the continuum 
calculation \cite{kura}. It is given in (\ref{meff}). After converting to the 
dimensionless 
integration variables $k_\mu$ of (\ref{z2sub}) one has for the part with 
the derivative acting on the denominator
\begin{eqnarray}
& &{\cal F}_{sub}^{den} =   \theta(\Lambda^2 - k^2)  \nl
& & \times \left\{
\frac{-4 \chi \, ( \chi^2 k_0^2 + b^2/4) \, ((k^2)^2 - b^2 \chi^2 k_0^2)}
{(k^2 + \lambda^2) \, ( (k^2)^2 + b^2 \chi^2 k_0^2)^2 }  
 + \,(\alpha_g - 1) \,\chi   
\frac{\chi^2 k_0^2 \, ( b^2 + 2 k^2)}{k^2\,(k^2 + \lambda^2)\,((k^2)^2 +
b^2 \chi^2k_0^2)}  \right\}, \nl
&& 
\end{eqnarray}
with
\be k^2 = \chi^2 k_0^2 + \sum_j k_j^2 \qquad \qquad b = 2 a_s m_{\it eff} .
\ee
The $\theta$-function imposes a cutoff on ${\cal F}_{sub}^{den}$ so 
that it vanishes identically for $k^2 > \Lambda^2$, where $\Lambda$ 
is some number $0 < \Lambda \leq \pi$. 
The subtraction term can be integrated analytically to give
\begin{eqnarray}
& &F^{den}(m_{\it eff} > 0,\Lambda,\lambda) = 
\int_k {\cal F}_{sub}^{den}(k,m_{\it eff}>0,\Lambda,\lambda)  \nl
& & = \frac{\alpha_s}{3 \pi} \left\{ \left[- 2 \ln(\frac{\Lambda^2}{\lambda^2})
\,+\, 2 \ln \left(\frac{\Lambda + \sqrt{b^2 + \Lambda^2}}{b}\right)
+ \frac{4 \Lambda^2}{b^4} \, ( b^2 + 3 \Lambda^2) + 
\frac{\sqrt{b^2 + \Lambda^2}}{b^4} 2 \Lambda (b^2 - 6 \Lambda^2 ) \right] 
\right.
\nl
 & & + (\alpha_g -1 ) \left. \left[\, \ln(\frac{\Lambda^2}{\lambda^2}) + 
\frac{2 \Lambda^2}{b^4} (\Lambda^2 + 2 b^2) -
\frac{ \Lambda ( 2 \Lambda^2 + 3 b^2)}{b^4}
\sqrt{b^2 + \Lambda^2} 
- \ln \left( \frac{\Lambda+\sqrt{b^2 + \Lambda^2}}{b}\right) \right]
\right\}  \nl
& &       
\end{eqnarray}
 and
\be
F^{den}(m_{\it eff}\equiv0,\Lambda,\lambda)  = 
 \frac{\alpha_s}{3 \pi} \,\left[- 1 + \frac{\alpha_g - 1}{2} \right] 
\ln(\frac{\Lambda^2}{\lambda^2}).
\ee
   The contribution in (\ref{dparts}) from 
 the derivative acting on the numerator is 
\begin{equation}
{\cal F}^{ num}_{ sub}
  =\theta(\Lambda^2-k^2)\left\{\frac{2\chi k^2}{((k^2)^2+b^2\chi^2k_0^2)
  (k^2+\lambda^2)}+(\alpha_g-1)\chi\frac{k^2-2\chi^2k_0^2}{((k^2)^2
+b^2\chi^2k_0^2)
  (k^2+\lambda^2)}\right\},
\end{equation}
which leads to 
\begin{eqnarray}
\lefteqn{F^{ num}(m_{\it eff}>0,\Lambda,\lambda)
  \ =\ \int_k{\cal F}^{ num}_{ sub}(k,m_{\it eff}>0,\Lambda,\lambda)}
  \nonumber\\
  &=&\frac{\alpha_s}{3\pi}\Bigg\{\left[4\ln\left(
  \frac{\Lambda+\sqrt{b^2+\Lambda^2}}b\right)-\frac{4\Lambda^2}{b^2}
  +\frac{4\Lambda}{b^2}\sqrt{b^2+\Lambda^2}\right]\nonumber\\&&
  +(\alpha_g-1)\left[\ln\left(\frac{\Lambda+\sqrt{b^2+\Lambda^2}}b\right)
  -\frac{2\Lambda^2}{b^4}(\Lambda^2+2b^2)
  +\frac{\Lambda(2\Lambda^2+3b^2)}{b^4}\sqrt{b^2+\Lambda^2}\right]\Bigg\}
\end{eqnarray}
and
\be
F^{num}(m_{\it eff}\equiv0,\Lambda,\lambda)  = 
 \frac{\alpha_s}{3 \pi} \,\left[\,  2 + \frac{\alpha_g - 1}{2} \right] 
\ln(\frac{\Lambda^2}{\lambda^2}) .
\ee
$F^{den} + F^{num}$ reproduces the IR divergent logarithms of 
(\ref{contir}).


\begin{table}
\label{tab:one}
\begin{center}
\begin{tabular}{ccc}
\multicolumn{1}{c}{action} &
\multicolumn{1}{c}{comments} &
\multicolumn{1}{c}{parameters} \\ \hline
 & & \\
$\act^A = \act^I_G + \act_{clover}$ &  massive and massless
  &  $c^G_0 = 5/3$, $c^G_1 = 
-1/12$, $\chi = 1$  \\
& & \\
$\act^{A^\prime} = \act^I_G + \act_{clover}$ &  massive and massless  & 
 $c^G_0 = 3.648$, $c^G_1 = -0.331$, $\chi = 1$ \\
 & & \\
$\act^B = \act^{II}_G + \act_{clover}$ &  massive and massless
  &   $\chi \geq 1$  \\
 & & \\
$\act^C = \act^{I}_G + \act^I_{D234}$ &   massless
  &  $c^G_0 = 5/3$, $c^G_1 = -1/12$, $\chi = 1$  \\
 & & \\
$\act^D = \act^{II}_G + \act^{II}_{D234}$ & massive and  massless
  &   $\chi \geq 1$  \\
 & & \\
\end{tabular}
\end{center}
\caption{Combinations of gauge and quark actions considered in this article.}
\end{table}
\begin{table}
\begin{center}
\begin{tabular}{c|rrrrrc}
 \multicolumn{7}{c}{ Action $\act^A$} \\\hline
 & regular& tadpole &reg + tad &t.i.&
$a_s M_{1,nosub}^{(1)}$ & $a_s M_{1,sub}^{(1)}$\\\hline
$a_s M_1^{(0)}$& \multicolumn{4}{c} {$\alpha_g = 1.0$} && \\\hline
$0.00$&$-1.770(3)$&$4.298$&$2.528(3)$&$-3.001$&$-0.472(3)$&$0.000\quad\;$\\
$0.01$&$-1.718(5)$&$4.266$&$2.548(5)$&$-2.978$&$-0.430(5)$&$0.037(6)$\\
$0.05$&$-1.547(3)$&$4.141$&$2.594(3)$&$-2.891$&$-0.297(3)$&$0.152(4)$\\
$0.10$&$-1.369(3)$&$3.991$&$2.623(3)$&$-2.786$&$-0.164(3)$&$0.263(4)$\\
$0.50$&$-0.463(3)$&$3.030$&$2.567(3)$&$-2.115$&$0.451(3)$&$0.737(3)$\\
$1.00$&$0.092(3)$&$2.260$&$2.352(3)$&$-1.578$&$0.774(3)$&$0.948(3)$\\
$2.00$&$0.530(3)$&$1.511$&$2.041(3)$&$-1.055$&$0.986(3)$&$1.050(3)$\\
$5.00$&$0.743(3)$&$1.096$&$1.839(3)$&$-0.765$&$1.074(3)$&$1.077(3)$\\
$10.00$&$0.752(3)$&$1.075$&$1.827(3)$&$-0.750$&$1.077(3)$&$1.077(3)$\\\hline
$a_s M_1^{(0)}$& \multicolumn{4}{c} {$\alpha_g = 0.0$} && \\\hline
$0.00$&$-0.472(3)$&$3.001$&$2.528(3)$&$-3.001$&$-0.472(3)$&$0.000\quad\;$\\
$0.01$&$-0.429(5)$&$2.978$&$2.549(5)$&$-2.978$&$-0.429(5)$&$0.038(6)$\\
$0.05$&$-0.293(3)$&$2.891$&$2.597(3)$&$-2.891$&$-0.293(3)$&$0.156(4)$\\
$0.10$&$-0.161(3)$&$2.786$&$2.626(3)$&$-2.786$&$-0.161(3)$&$0.266(4)$\\
$0.50$&$0.454(3)$&$2.115$&$2.569(3)$&$-2.115$&$0.454(3)$&$0.740(3)$\\
$1.00$&$0.775(3)$&$1.578$&$2.353(3)$&$-1.578$&$0.775(3)$&$0.949(3)$\\
$2.00$&$0.987(3)$&$1.055$&$2.042(3)$&$-1.055$&$0.987(3)$&$1.051(3)$\\
$5.00$&$1.076(3)$&$0.765$&$1.841(3)$&$-0.765$&$1.076(3)$&$1.079(3)$\\
$10.00$&$1.079(3)$&$0.750$&$1.829(3)$&$-0.750$&$1.079(3)$&$1.079(3)$\\
\end{tabular}
\caption{One-loop mass renormalization for action ${\cal S}^A$ 
for different  $a_sM_1^{(0)}$ values.  Results are presented for 
two choices of the gauge parameter $\alpha_g = 1$ and $\alpha_g = 0$.
$a_s M_{1,nosub}^{(1)}$ is the same as (reg + tad + t.i.).  
$a_sM_{1,sub}^{(1)}$ is defined in ~\protect(\ref{m11b}) and 
related to $a_sM_{1,nosub}^{(1)}$ in ~\protect(\ref{m1sub}).  Where 
errors are not indicated explicitly, they are of $O(1)$ or less 
in the last digit.
}
\end{center}
\end{table}
\begin{table}
\begin{center}
\begin{tabular}{c|crc|crc}
& \multicolumn{3}{c|} { Action ${\cal S}^{A^\prime}$} & 
\multicolumn{3}{c}{Action ${\cal S}^C$} \\\hline
$a_s M_1^{(0)}$& reg + tad & $a_sM_{1,nosub}^{(1)}$&
 $a_s M_{1,sub}^{(1)} \quad$
& reg + tad &$a_sM_{1,nosub}^{(1)}$& $a_s M_{1,sub}^{(1)} \quad$ \\\hline
$0.00$&$1.480$&$-0.397 \quad$&$0.000$ &$2.213$ &$-1.287\quad$ &$0.000$  \\
$0.01$&$1.505$&$-0.358 \quad$&$0.035$ & & & \\
$0.05$&$1.566$&$-0.242 \quad$&$0.136$ & & &  \\
$0.10$&$1.612$&$-0.131 \quad$&$0.228$ & & &  \\
$0.50$&$1.690$&$0.367 \quad$& $0.608$ & & &  \\
$1.00$&$1.611$&$0.624 \quad$&$0.770$ & & &  \\
$2.00$&$1.466$&$0.806 \quad$&$0.859$ & & &  \\
$5.00$&$1.376$&$0.897 \quad$&$0.900$ & & &  \\
$10.00$&$1.371$&$0.901 \quad$&$0.901$ & & &  \\
\end{tabular}
\caption{One-loop mass renormalization for actions ${\cal S}^{A^\prime}$ 
and ${\cal S}^C$.  
Numerical integration errors are at the $\pm0.006$ level
 for $a_sM_1^{(0)} =0.01$
and of $O(4)$ or less in the last digit otherwise.
}
\end{center}
\end{table}
\begin{table}
\begin{center}
\begin{tabular}{c|crc|crc}
 \multicolumn{7}{c}{ Action $\act^B$} \\\hline
& reg + tad & $a_sM_{1,nosub}^{(1)}$&
 $a_s M_{1,sub}^{(1)} \quad$
& reg + tad &$a_sM_{1,nosub}^{(1)}$& $a_s M_{1,sub}^{(1)} \quad$ \\\hline
$a_sM_1^{(0)}$ & 
\multicolumn{3}{c|}{$\chi = 1.0 \qquad$} &
\multicolumn{3}{c}{$\chi = 2.0 \qquad$} \\\hline
$0.00$&$2.710$&$-0.479\quad$&$0.000$&$3.240$&$-0.881\quad$&$0.000$  \\
$0.01$&$2.729$&$-0.436 \quad$&$0.038$&$3.250$&$-0.833 \quad$&$0.039$ \\
$0.05$&$2.774$&$-0.299 \quad$&$0.157$&$3.260$&$-0.675 \quad$&$0.164$  \\
$0.10$&$2.799$&$-0.164 \quad$&$0.269$&$3.251$&$-0.512 \quad$&$0.288$  \\
$0.50$&$2.717$&$0.463 \quad$&$0.754$&$3.060$&$0.290 \quad$&$0.863$  \\
$1.00$&$2.466$&$0.779 \quad$&$0.955$&$2.824$&$0.772 \quad$&$1.178$  \\
$2.00$&$2.096$&$0.961 \quad$&$1.026$&$2.461$&$1.155 \quad$&$1.381$  \\
$5.00$&$1.842$&$1.014 \quad$&$1.017$&$1.978$&$1.371 \quad$&$1.419$  \\\hline
$a_sM_1^{(0)}$ & 
\multicolumn{3}{c|}{$\chi = 3.0 \qquad$} &
\multicolumn{3}{c}{$\chi = 3.6 \qquad$} \\\hline
$0.00$&$3.298$&$-1.139 \quad$&$0.000$&$3.286$&$-1.243 \quad$&$0.000$  \\
$0.01$&$3.302$&$-1.092 \quad$&$0.036$&$3.289$&$-1.195 \quad$&$0.036$ \\
$0.05$&$3.312$&$-0.918 \quad$&$0.167$&$3.299$&$-1.017 \quad$&$0.167$  \\
$0.10$&$3.299$&$-0.742 \quad$&$0.293$&$3.287$&$-0.835 \quad$&$0.295$  \\
$0.50$&$3.110$&$0.145 \quad$&$0.896$&$3.106$&$0.083 \quad$&$0.905$  \\
$1.00$&$2.901$&$0.695 \quad$&$1.244$&$2.910$&$0.656 \quad$&$1.262$  \\
$2.00$&$2,589$&$1.161 \quad$&$1.498$&$2.620$&$1.151 \quad$&$1.532$  \\
$5.00$&$2.106$&$1.472 \quad$&$1.582$&$2.158$&$1.496 \quad$&$1.636$  \\\hline
$a_sM_1^{(0)}$ & 
\multicolumn{3}{c|}{$\chi = 4.0 \qquad$} &
\multicolumn{3}{c}{$\chi = 5.0 \qquad$} \\\hline
$0.00$&$3.271$&$-1.299 \quad$&$0.000$&$3.233$&$-1.405 \quad$&$0.000$  \\
$0.01$&$3.271$&$-1.255 \quad$&$0.031$&$3.233$&$-1.360 \quad$&$0.031$ \\
$0.05$&$3.285$&$-1.070 \quad$&$0.167$&$3.249$&$-1.170 \quad$&$0.168$  \\
$0.10$&$3.274$&$-0.885 \quad$&$0.296$&$3.239$&$-0.980 \quad$&$0.297$  \\
$0.50$&$3.099$&$0.049 \quad$&$0.910$&$3.076$&$-0.018 \quad$&$0.915$  \\
$1.00$&$2.911$&$0.634 \quad$&$1.270$&$2.903$&$0.590 \quad$&$1.283$  \\
$2.00$&$2.631$&$1.143 \quad$&$1.547$&$2.645$&$1.124 \quad$&$1.572$  \\
$5.00$&$2.184$&$1.505 \quad$&$1.662$&$2.228$&$1.518 \quad$&$1.707$  \\\hline
$a_sM_1^{(0)}$ & 
\multicolumn{3}{c|}{$\chi = 5.3 \qquad$} &
\multicolumn{3}{c}{$\chi = 6.0 \qquad$} \\\hline
$0.00$&$3.222$&$-1.429 \quad$&$0.000$&$3.197$&$-1.479 \quad$&$0.000$  \\
$0.01$&$3.226$&$-1.380 \quad$&$0.035$&$3.204$&$-1.426 \quad$&$0.038$ \\
$0.05$&$3.238$&$-1.194 \quad$&$0.167$&$3.213$&$-1.241 \quad$&$0.168$  \\
$0.10$&$3.228$&$-1.003 \quad$&$0.296$&$3.205$&$-1.048 \quad$&$0.296$  \\
$0.50$&$3.069$&$-0.034 \quad$&$0.915$&$3.053$&$-0.065 \quad$&$0.918$  \\
$1.00$&$2.899$&$0.579 \quad$&$1.285$&$2.891$&$0.557 \quad$&$1.290$  \\
$2.00$&$2.647$&$1.119 \quad$&$1.577$&$2.649$&$1.108 \quad$&$1.586$  \\
$5.00$&$2.237$&$1.520 \quad$&$1.717$&$2.254$&$1.522 \quad$&$1.734$  \\
\end{tabular}
\caption{One-loop mass renormalization for action ${\cal S}^B$ 
for several values of the anisotropy $ \chi = a_s/a_t$. 
Numerical integration errors are as in Table III.
}
\end{center}
\end{table}
\begin{table}
\begin{center}
\begin{tabular}{c|crc|crc}
 \multicolumn{7}{c}{ Action $\act^D$} \\\hline
& reg + tad & $a_sM_{1,nosub}^{(1)}$&
 $a_s M_{1,sub}^{(1)} \quad$
& reg + tad &$a_sM_{1,nosub}^{(1)}$& $a_s M_{1,sub}^{(1)} \quad$ \\\hline
$a_sM_1^{(0)}$ & 
\multicolumn{3}{c|}{$\chi = 1.0 \qquad$} &
\multicolumn{3}{c}{$\chi = 2.0 \qquad$} \\\hline
$0.00$&$2.247$&$-1.339 \quad$&$0.000$&$2.740$&$-1.927 \quad$&$0.000$  \\
$0.01$&$2.268$&$-1.289 \quad$&$0.037$&$2.744$&$-1.878 \quad$&$0.030$ \\
$0.05$&$2.329$&$-1.121 \quad$&$0.153$&$2.776$&$-1.678 \quad$&$0.157$  \\
$0.10$&$2.375$&$-0.947 \quad$&$0.265$&$2.785$&$-1.473 \quad$&$0.277$  \\
$0.50$&$2.451$&$-0.043 \quad$&$0.769$&$2.728$&$-0.396 \quad$&$0.858$  \\
$1.00$&$2.342$&$ 0.510 \quad$&$1.003$&$2.623$&$ 0.320 \quad$&$1.208$ \\
$2.00$&$2.082$&$ 0.894 \quad$&$1.075$&$2.404$&$ 0.958 \quad$&$1.453$  \\
$5.00$&$1.843$&$ 1.012 \quad$&$1.021$&$1.986$&$ 1.349 \quad$&$1.455$  \\\hline
$a_sM_1^{(0)}$ & 
\multicolumn{3}{c|}{$\chi = 3.0 \qquad$} &
\multicolumn{3}{c}{$\chi = 3.6 \qquad$} \\\hline
$0.00$&$2.788$&$-2.250 \quad$&$0.000$&$2,777$&$-2.370 \quad$&$0.000$  \\
$0.01$&$2.794$&$-2.195 \quad$&$0.033$&$2.781$&$-2.315 \quad$&$0.032$ \\
$0.05$&$2.818$&$-1.984 \quad$&$0.159$&$2.805$&$-2.099 \quad$&$0.158$  \\
$0.10$&$2.824$&$-1.762 \quad$&$0.282$&$2.811$&$-1.872 \quad$&$0.282$  \\
$0.50$&$2.759$&$-0.601 \quad$&$0.883$&$2.751$&$-0.680 \quad$&$0.888$  \\
$1.00$&$2.674$&$ 0.178 \quad$&$1.262$&$2.675$&$ 0.120 \quad$&$1.275$  \\
$2.00$&$2.506$&$ 0.901 \quad$&$1.566$&$2.528$&$ 0.870 \quad$&$1.596$  \\
$5.00$&$2.115$&$ 1.423 \quad$&$1.639$&$2.166$&$ 1.435 \quad$&$1.701$  \\\hline
$a_sM_1^{(0)}$ & 
\multicolumn{3}{c|}{$\chi = 4.0 \qquad$} &
\multicolumn{3}{c}{$\chi = 5.0 \qquad$} \\\hline
$0.00$&$2.769$&$-2.427 \quad$&$0.000$&$2.732$&$-2.544 \quad$&$0.000$  \\
$0.01$&$2.768$&$-2.377 \quad$&$0.026$&$2.734$&$-2.490 \quad$&$0.029$ \\
$0.05$&$2.792$&$-2.152 \quad$&$0.159$&$2.760$&$-2.267 \quad$&$0.156$  \\
$0.10$&$2.799$&$-1.928 \quad$&$0.278$&$2.768$&$-2.030 \quad$&$0.282$  \\
$0.50$&$2.742$&$-0.722 \quad$&$0.886$&$2.719$&$-0.798 \quad$&$0.891$  \\
$1.00$&$2.672$&$ 0.090 \quad$&$1.279$&$2.660$&$ 0.033 \quad$&$1.288$  \\
$2.00$&$2.536$&$ 0.853 \quad$&$1.609$&$2.542$&$ 0.817 \quad$&$1.629$  \\
$5.00$&$2.190$&$ 1.436 \quad$&$1.729$&$2.232$&$ 1.436 \quad$&$1.779$  \\\hline
$a_sM_1^{(0)}$ & 
\multicolumn{3}{c|}{$\chi = 5.3 \qquad$} &
\multicolumn{3}{c}{$\chi = 6.0 \qquad$} \\\hline
$0.00$&$2.720$&$-2.572 \quad$&$0.000$&$2.701$&$-2.620 \quad$&$0.000$  \\
$0.01$&$2.724$&$-2.516 \quad$&$0.031$&$2.708$&$-2.561 \quad$&$0.033$ \\
$0.05$&$2.759$&$-2.283 \quad$&$0.166$&$2.732$&$-2.337 \quad$&$0.158$  \\
$0.10$&$2.765$&$-2.047 \quad$&$0.291$&$2.740$&$-2.099 \quad$&$0.283$  \\
$0.50$&$2.717$&$-0.810 \quad$&$0.899$&$2.699$&$-0.848 \quad$&$0.894$  \\
$1.00$&$2.658$&$ 0.022 \quad$&$1.293$&$2.646$&$-0.007 \quad$&$1.291$  \\
$2.00$&$2.543$&$ 0.809 \quad$&$1.633$&$2.541$&$ 0.791 \quad$&$1.638$  \\
$5.00$&$2.242$&$ 1.436 \quad$&$1.791$&$2.256$&$ 1.432 \quad$&$1.808$  \\
\end{tabular}
\caption{One-loop mass renormalization for action ${\cal S}^D$ 
for several values of the anisotropy $ \chi = a_s/a_t$. 
Numerical integration errors are as in Table III.
}
\end{center}
\end{table}
\begin{table}
\begin{center}
\begin{tabular}{c|rrr|rrr}
 \multicolumn{7}{c}{ Action $\act^A$} \\\hline
 & \multicolumn{3}{c|}{$\alpha_g = 1.0 \qquad$} &
\multicolumn{3}{c}{$\alpha_g = 0.0 \qquad$} \\\hline
$a_s M_1^{(0)}$& 
\multicolumn{1}{c|}
{$Z_{2,dp_0}^{(1)}$ no $t.i.$ }& 
\multicolumn{1}{c|}
{$Z_{2,dp_0}^{(1)}$ with $t.i.$} &
\multicolumn{1}{c|}
{ $Z_2^{(1)}$ with $t.i.$} &
\multicolumn{1}{c|}
{$Z_{2,dp_0}^{(1)}$ no $t.i.$ }& 
\multicolumn{1}{c|}
{$Z_{2,dp_0}^{(1)}$ with $t.i.$ }&
\multicolumn{1}{c}
{ $Z_2^{(1)}$ with $t.i.$}
\\\hline
$0.00$&$0.687(3) \quad$&$-0.063(3)\quad$&$-0.063(3)\quad$&
$1.195(3) \quad$&$0.445(3) \quad$&$0.445(3) \quad$\\
$0.01$&$-2.688(6)\quad$&$-3.438(6)\quad$&$-3.475(7)\quad$&
$-2.169(6) \quad$&$-2.919(6) \quad$&$-2.956(7) \quad$\\
$0.05$&$-1.569(5)\quad$&$-2.319(5)\quad$&$-2.471(6)\quad$&
$-1.052(5) \quad$&$-1.802(5) \quad$&$-1.954(6) \quad$\\
$0.10$&$-1.041(5)\quad$&$-1.791(5)\quad$&$-2.054(6)\quad$&
$-0.524(5) \quad$&$-1.274(5) \quad$&$-1.537(6) \quad$\\
$0.50$&$0.393(3)\quad$&$-0.357(3)\quad$&$-1.094(4)\quad$&
$0.905(3) \quad$&$0.155(3) \quad$&$-0.582(4) \quad$\\
$1.00$&$1.112(3)\quad$&$0.362(3)\quad$&$-0.586(4)\quad$&
$1.623(3) \quad$&$0.873(3) \quad$&$-0.075(4) \quad$\\
$2.00$&$1.795(3)\quad$&$1.045(3)\quad$&$-0.005(4)\quad$&
$2.304(3) \quad$&$1.554(3) \quad$&$0.504(4) \quad$\\
$5.00$&$2.223(20)\;\;$&$1.473(20)\;\;$&$0.396(20)\;\;$&
$2.719(20)\;\;$&$1.969(20)\;\;$&$0.892(20)\;\;$\\
\end{tabular}
\caption{One-loop wave function renormalization for action ${\cal S}^A$ 
for different  $a_sM_1^{(0)}$ values.  Results are presented for 
two choices for the gauge parameter $\alpha_g = 1$ and $\alpha_g = 0$.
$Z_2^{(1)}$ includes both $Z_{2,dp_0}^{(1)}$ and $Z_{2,M_1}^{(1)}$
as defined in equation ~\protect(\ref{z2one}).  
}
\end{center}
\end{table}
\begin{table}
\begin{center}
\begin{tabular}{c|rr}
 \multicolumn{3}{c}{Action ${\cal S}^A \qquad \alpha_g = 1.0 $} \\\hline
$a_s M_1^{(0)}$& 
\multicolumn{1}{c}
{$Z_{2,diff}^{(1)}$ no $t.i. $ }& 
\multicolumn{1}{c}
{$Z_{2,diff}^{(1)}$ with $t.i. \qquad$} \\\hline
$0.00$&$0.634(3) \qquad$&$-0.116(3)\qquad \qquad$\\
$0.01$&$0.668(6)\qquad$&$-0.082(6)\qquad \qquad$\\
$0.05$&$0.763(5)\qquad$&$0.013(5)\qquad \qquad$\\
$0.10$&$0.849(5)\qquad$&$0.099(5)\qquad \qquad$\\
$0.50$&$1.259(3)\qquad$&$0.509(3)\qquad \qquad$\\
$1.00$&$1.536(3)\qquad$&$0.786(3)\qquad \qquad$\\
$2.00$&$1.778(3)\qquad$&$1.028(3)\qquad \qquad$\\
$5.00$&$1.623(20)\;\;\quad$&$0.873(20)\;\;\quad \qquad$\\
\end{tabular}
\caption{$Z_{2,dp_0}^{(1)}$ with $\ln(am)$ contributions subtracted out
 for action ${\cal S}^A$ in Feynman gauge.
$Z_{2,diff}^{(1)}$ is defined  in equation ~\protect(\ref{z2diff}).  
}
\end{center}
\end{table}
\begin{table}
\begin{center}
\begin{tabular}{c|rrr|rrr}
 \multicolumn{7}{c}{ Action $\act^{A^\prime} $} \\\hline
 & \multicolumn{3}{c|}{$\alpha_g = 1.0 \qquad$} &
\multicolumn{3}{c}{$\alpha_g = 0.0 \qquad$} \\\hline
$a_s M_1^{(0)}$& 
\multicolumn{1}{c|}
{$Z_{2,dp_0}^{(1)}$ no $t.i.$ }& 
\multicolumn{1}{c|}
{$Z_{2,dp_0}^{(1)}$ with $t.i.$} &
\multicolumn{1}{c|}
{ $Z_2^{(1)}$ with $t.i.$} &
\multicolumn{1}{c|}
{$Z_{2,dp_0}^{(1)}$ no $t.i.$ }& 
\multicolumn{1}{c|}
{$Z_{2,dp_0}^{(1)}$ with $t.i.$ }&
\multicolumn{1}{c}
{ $Z_2^{(1)}$ with $t.i.$}
\\\hline
$0.00$&$0.239\quad$&$-0.230\quad$&$-0.230\quad$&$0.747\quad$&$0.278\quad$&$
0.278\quad$\\
$0.01$&$-3.137\quad$&$-3.606\quad$&$-3.641\quad$&$-2.628\quad$&$-3.097\quad$&$
-3.132\quad$\\
$0.05$&$-2.024\quad$&$-2.493\quad$&$-2.629\quad$&$-1.515\quad$&$-1.984\quad
$&$-2.120\quad$\\
$0.10$&$-1.502\quad$&$-1.971\quad$&$-2.199\quad$&$-0.993\quad$&$-1.462\quad
$&$-1.690\quad$\\
$0.50$&$-0.120\quad$&$-0.590\quad$&$-1.198\quad$&$0.389\quad$&$-0.081\quad
$&$-0.689\quad$\\
$1.00$&$0.546\quad$&$0.076\quad$&$-0.694\quad$&$1.054\quad$&$0.585\quad
$&$-0.185\quad$\\
$2.00$&$1.165\quad$&$0.695\quad$&$-0.164\quad$&$1.674\quad$&$1.204\quad
$&$0.345\quad$\\
$5.00$&$1.568\quad$&$1.099\quad$&$0.199\quad$&$2.079\quad$&$1.610\quad
$&$0.710\quad$
\\\hline\hline
 \multicolumn{7}{c}{ Action $\act^C $} \\\hline
 & \multicolumn{3}{c|}{$\alpha_g = 1.0 \qquad$} &
\multicolumn{3}{c}{$\alpha_g = 0.0 \qquad$} \\\hline
$a_s M_1^{(0)}$& 
\multicolumn{1}{c|}
{$Z_{2,dp_0}^{(1)}$ no $t.i.$ }& 
\multicolumn{1}{c|}
{$Z_{2,dp_0}^{(1)}$ with $t.i.$} &
\multicolumn{1}{c|}
{ $Z_2^{(1)}$ with $t.i.$} &
\multicolumn{1}{c|}
{$Z_{2,dp_0}^{(1)}$ no $t.i.$ }& 
\multicolumn{1}{c|}
{$Z_{2,dp_0}^{(1)}$ with $t.i.$ }&
\multicolumn{1}{c}
{ $Z_2^{(1)}$ with $t.i.$}
\\\hline
$0.00$&$0.145\quad$&$-0.355\quad$&$-0.355\quad$&$0.653\quad$&
$0.153\quad$&$0.153\quad$\\
\end{tabular}
\caption{One-loop wave function renormalization for actions
 ${\cal S}^{A^\prime}$ and ${\cal S}^C$. 
Numerical integration errors are at the $\pm0.02$ level for 
$a_sM_1^{(0)} = 5.0$, at the $\pm0.006$ level 
for $a_sM_1^{(0)}$ = 0.01, 0.05 and 0.10 
 and of $O(4)$ in the last digit for other masses.
}
\end{center}
\end{table}
\begin{table}
\begin{center}
\begin{tabular}{c|rrr|rrr}
 \multicolumn{7}{c}{ Action $\act^B \qquad \alpha_g = 1.0$} \\\hline
& \multicolumn{1}{c|}
{$Z_{2,dp_0}^{(1)}$ no $t.i.$ }& 
\multicolumn{1}{c|}
{$Z_{2,dp_0}^{(1)}$ with $t.i.$} &
\multicolumn{1}{c|}
{ $Z_2^{(1)}$ with $t.i.$} &
\multicolumn{1}{c|}
{$Z_{2,dp_0}^{(1)}$ no $t.i.$ }& 
\multicolumn{1}{c|}
{$Z_{2,dp_0}^{(1)}$ with $t.i.$ }&
\multicolumn{1}{c}
{ $Z_2^{(1)}$ with $t.i.$}
\\\hline
$a_sM_1^{(0)}$ & 
\multicolumn{3}{c|}{$\chi = 1.0 \qquad$} &
\multicolumn{3}{c}{$\chi = 2.0 \qquad$} \\\hline
$0.00$&$0.789\quad$&$-0.023\quad$&$-0.023\quad$&$0.323\quad$&$0.110\quad$&$
0.110\quad$  \\
$0.01$&$-2.586\quad$&$-3.398\quad$&$-3.431\quad$&$-3.050\quad$&$-3.263\quad
$&$-3.293\quad$  \\
$0.05$&$-1.469\quad$&$-2.281\quad$&$-2.437\quad$&$-1.930\quad$&$-2.143\quad
$&$-2.301\quad$ \\
$0.10$&$-0.942\quad$&$-1.755\quad$&$-2.024\quad$&$-1.406\quad$&$-1.619\quad
$&$-1.887\quad$  \\
$0.50$&$0.467\quad$&$-0.345\quad$&$-1.099\quad$&$-0.111\quad$&$-0.324\quad
$&$-0.974\quad$  \\
$1.00$&$1.150\quad$&$0.337\quad$&$-0.618\quad$&$0.398\quad$&$0.185\quad
$&$-0.569\quad$  \\
$2.00$&$1.766\quad$&$0.953\quad$&$-0.073\quad$&$0.822\quad$&$0.609\quad
$&$-0.147\quad$  \\
$5.00$&$2.124\quad$&$1.312\quad$&$0.295\quad$&$1.237\quad$&$1.025\quad
$&$0.312\quad$  \\\hline
$a_sM_1^{(0)}$ & 
\multicolumn{3}{c|}{$\chi = 3.0 \qquad$} &
\multicolumn{3}{c}{$\chi = 3.6 \qquad$} \\\hline
$0.00$&$0.265\quad$&$0.172\quad$&$0.172\quad$&$0.253\quad$&$0.190\quad$&$0.190\quad$  \\
$0.01$&$-3.107\quad$&$-3.200\quad$&$-3.231\quad$&$-3.118\quad$&$-3.182\quad
$&$-3.213\quad$ \\
$0.05$&$-1.988\quad$&$-2.080\quad$&$-2.240\quad$&$-1.999\quad$&$-2.062\quad
$&$-2.222\quad$  \\
$0.10$&$-1.465\quad$&$-1.558\quad$&$-1.828\quad$&$-1.476\quad$&$-1.540\quad
$&$-1.810\quad$  \\
$0.50$&$-0.198\quad$&$-0.290\quad$&$-0.923\quad$&$-0.216\quad$&$-0.279\quad
$&$-0.907\quad$  \\
$1.00$&$0.259\quad$&$0.167\quad$&$-0.534\quad$&$0.228\quad$&$0.165\quad
$&$-0.522\quad$  \\
$2.00$&$0.594\quad$&$0.501\quad$&$-0.150\quad$&$0.536\quad$&$0.473\quad
$&$-0.147\quad$  \\
$5.00$&$0.911\quad$&$0.819\quad$&$0.272\quad$&$0.807\quad$&$0.744\quad
$&$0.256\quad$  \\\hline
$a_sM_1^{(0)}$ & 
\multicolumn{3}{c|}{$\chi = 4.0 \qquad$} &
\multicolumn{3}{c}{$\chi = 5.0 \qquad$} \\\hline
$0.00$&$0.249\quad$&$0.198\quad$&$0.198\quad$&$0.242\quad$&$0.210\quad
$&$0.210\quad$  \\
$0.01$&$-3.123\quad$&$-3.174\quad$&$-3.205\quad$&$-3.130\quad$&$-3.162\quad
$&$-3.193\quad$ \\
$0.05$&$-2.003\quad$&$-2.054\quad$&$-2.214\quad$&$-2.010\quad$&$-2.042\quad
$&$-2.202\quad$  \\
$0.10$&$-1.481\quad$&$-1.532\quad$&$-1.803\quad$&$-1.487\quad$&$-1.519\quad
$&$-1.790\quad$  \\
$0.50$&$-0.223\quad$&$-0.275\quad$&$-0.901\quad$&$-0.233\quad$&$-0.266\quad
$&$-0.889\quad$  \\
$1.00$&$0.215\quad$&$0.164\quad$&$-0.517\quad$&$0.196\quad$&$0.164\quad
$&$-0.508\quad$  \\
$2.00$&$0.511\quad$&$0.460\quad$&$-0.146\quad$&$0.473\quad$&$0.441\quad
$&$-0.143\quad$  \\
$5.00$&$0.758\quad$&$0.707\quad$&$0.249\quad$&$0.678\quad$&$0.645\quad
$&$0.236\quad$ \\\hline
$a_sM_1^{(0)}$ & 
\multicolumn{3}{c|}{$\chi = 5.3 \qquad$} &
\multicolumn{3}{c}{$\chi = 6.0 \qquad$} \\\hline
$0.00$&$0.241\quad$&$0.212\quad$&$0.212\quad$&$0.238\quad$&$0.216\quad
$&$0.216\quad$  \\
$0.01$&$-3.131\quad$&$-3.160\quad$&$-3.191\quad$&$-3.133\quad$&$-3.156\quad
$&$-3.186\quad$ \\
$0.05$&$-2.011\quad$&$-2.040\quad$&$-2.200\quad$&$-2.013\quad$&$-2.036\quad
$&$-2.196\quad$  \\
$0.10$&$-1.488\quad$&$-1.517\quad$&$-1.788\quad$&$-1.490\quad$&$-1.513\quad
$&$-1.783\quad$  \\
$0.50$&$-0.236\quad$&$-0.264\quad$&$-0.887\quad$&$-0.239\quad$&$-0.261\quad
$&$-0.882\quad$  \\
$1.00$&$0.193\quad$&$0.164\quad$&$-0.506\quad$&$0.186\quad$&$0.164\quad
$&$-0.502\quad$  \\
$2.00$&$0.466\quad$&$0.437\quad$&$-0.142\quad$&$0.453\quad$&$0.431\quad
$&$-0.140\quad$  \\
$5.00$&$0.661\quad$&$0.633\quad$&$0.234\quad$&$0.631\quad$&$0.609\quad
$&$0.229\quad$  \\
\end{tabular}
\caption{One-loop wave function renormalization for action ${\cal S}^D$ 
in Feynman gauge.
Numerical integration errors are at the $\pm0.02$ level for 
$a_sM_1^{(0)} = 5.0$ and $\chi < 3$, at the $\pm0.006$ level 
for $a_sM_1^{(0)}$ = 0.01, 0.05 and 0.10 
 and of $O(4)$ in the last digit for all other cases.
}
\end{center}
\end{table}
\begin{table}
\begin{center}
\begin{tabular}{c|rrr|rrr}
 \multicolumn{7}{c}{ Action $\act^B \qquad \alpha_g = 0.0$} \\\hline
& \multicolumn{1}{c|}
{$Z_{2,dp_0}^{(1)}$ no $t.i.$ }& 
\multicolumn{1}{c|}
{$Z_{2,dp_0}^{(1)}$ with $t.i.$} &
\multicolumn{1}{c|}
{ $Z_2^{(1)}$ with $t.i.$} &
\multicolumn{1}{c|}
{$Z_{2,dp_0}^{(1)}$ no $t.i.$ }& 
\multicolumn{1}{c|}
{$Z_{2,dp_0}^{(1)}$ with $t.i.$ }&
\multicolumn{1}{c}
{ $Z_2^{(1)}$ with $t.i.$}
\\\hline
$a_sM_1^{(0)}$ & 
\multicolumn{3}{c|}{$\chi = 1.0 \qquad$} &
\multicolumn{3}{c}{$\chi = 2.0 \qquad$} \\\hline
$0.00$&$1.297\quad$&$0.485\quad$&$0.485\quad$&$0.833\quad$&$0.620\quad
$&$0.620\quad$  \\
$0.01$&$-2.077\quad$&$-2.889\quad$&$-2.930\quad$&$-2.539\quad$&$-2.751\quad
$&$-2.793\quad$ \\
$0.05$&$-0.960\quad$&$-1.772\quad$&$-1.931\quad$&$-1.419\quad$&$-1.632\quad
$&$-1.794\quad$  \\
$0.10$&$-0.433\quad$&$-1.246\quad$&$-1.518\quad$&$-0.895\quad$&$-1.108
\quad$&$-1.379\quad$  \\
$0.50$&$0.975\quad$&$0.163\quad$&$-0.593\quad$&$0.400\quad$&$0.187\quad
$&$-0.465\quad$  \\
$1.00$&$1.658\quad$&$0.845\quad$&$-0.111\quad$&$0.908\quad$&$0.696\quad
$&$-0.058\quad$  \\
$2.00$&$2.274\quad$&$1.461\quad$&$0.434\quad$&$1.332\quad$&$1.119\quad
$&$0.363\quad$  \\
$5.00$&$2.635\quad$&$1.823\quad$&$0.804\quad$&$1.748\quad$&$1.535\quad
$&$0.822\quad$ \\\hline
$a_sM_1^{(0)}$ & 
\multicolumn{3}{c|}{$\chi = 3.0 \qquad$} &
\multicolumn{3}{c}{$\chi = 3.6 \qquad$} \\\hline
$0.00$&$ 0.763\quad$&$ 0.670\quad$&$ 0.670\quad$&$ 0.747\quad$&$ 0.683\quad$&$
 0.683\quad$  \\
$0.01$&$-2.608\quad$&$-2.701\quad$&$-2.742\quad$&$-2.624\quad$&$-2.688\quad$&$
-2.728\quad$ \\
$0.05$&$-1.489\quad$&$-1.581\quad$&$-1.745\quad$&$-1.505\quad$&$-1.568\quad$&$
-1.732\quad$  \\
$0.10$&$-0.966\quad$&$-1.059\quad$&$-1.332\quad$&$-0.982\quad$&$-1.045\quad$&$
-1.319\quad$  \\
$0.50$&$ 0.301\quad$&$ 0.208\quad$&$-0.426\quad$&$ 0.279\quad$&$ 0.215\quad$&$
-0.415\quad$  \\
$1.00$&$ 0.757\quad$&$ 0.665\quad$&$-0.037\quad$&$ 0.722\quad$&$ 0.658\quad$&$
-0.029\quad$  \\
$2.00$&$ 1.092\quad$&$ 0.999\quad$&$ 0.348\quad$&$ 1.029\quad$&$ 0.966\quad$&$
 0.346\quad$  \\
$5.00$&$ 1.408\quad$&$ 1.316\quad$&$ 0.770\quad$&$ 1.301\quad$&$ 1.237\quad$&$
 0.750\quad$  \\\hline
$a_sM_1^{(0)}$ & 
\multicolumn{3}{c|}{$\chi = 4.0 \qquad$} &
\multicolumn{3}{c}{$\chi = 5.0 \qquad$} \\\hline
$0.00$&$ 0.740\quad$&$ 0.689\quad$&$ 0.689\quad$&$ 0.729\quad$&$ 0.697\quad$&$
 0.697\quad$  \\
$0.01$&$-2.631\quad$&$-2.682\quad$&$-2.722\quad$&$-2.642\quad$&$-2.674\quad$&$
-2.713\quad$ \\
$0.05$&$-1.511\quad$&$-1.562\quad$&$-1.726\quad$&$-1.522\quad$&$-1.554\quad$&$
-1.718\quad$  \\
$0.10$&$-0.988\quad$&$-1.040\quad$&$-1.313\quad$&$-0.999\quad$&$-1.031\quad$&$
-1.305\quad$  \\
$0.50$&$ 0.269\quad$&$ 0.217\quad$&$-0.410\quad$&$ 0.255\quad$&$ 0.222\quad$&$
-0.403\quad$  \\
$1.00$&$ 0.706\quad$&$ 0.655\quad$&$-0.026\quad$&$ 0.684\quad$&$ 0.651\quad$&$
-0.021\quad$  \\
$2.00$&$ 1.002\quad$&$ 0.951\quad$&$ 0.345\quad$&$ 0.960\quad$&$ 0.928\quad$&$
 0.345\quad$  \\
$5.00$&$ 1.249\quad$&$ 1.198\quad$&$ 0.740\quad$&$ 1.165\quad$&$ 1.133\quad$&$
 0.724\quad$  \\\hline
$a_sM_1^{(0)}$ & 
\multicolumn{3}{c|}{$\chi = 5.3 \qquad$} &
\multicolumn{3}{c}{$\chi = 6.0 \qquad$} \\\hline
$0.00$&$ 0.727\quad$&$ 0.698\quad$&$ 0.698\quad$&$ 0.723\quad$&$ 0.701\quad$&$
 0.701\quad$  \\
$0.01$&$-2.644\quad$&$-2.673\quad$&$-2.712\quad$&$-2.648\quad$&$-2.670\quad$&$
-2.709\quad$ \\
$0.05$&$-1.524\quad$&$-1.553\quad$&$-1.716\quad$&$-1.528\quad$&$-1.550\quad$&$
-1.714\quad$  \\
$0.10$&$-1.001\quad$&$-1.030\quad$&$-1.303\quad$&$-1.005\quad$&$-1.027\quad$&$
-1.301\quad$  \\
$0.50$&$ 0.252\quad$&$ 0.223\quad$&$-0.401\quad$&$ 0.247\quad$&$ 0.225\quad$&$
-0.398\quad$  \\
$1.00$&$ 0.679\quad$&$ 0.650\quad$&$-0.020\quad$&$ 0.672\quad$&$ 0.649\quad$&$
-0.017\quad$  \\
$2.00$&$ 0.952\quad$&$ 0.923\quad$&$ 0.344\quad$&$ 0.937\quad$&$ 0.915\quad$&$
 0.345\quad$  \\
$5.00$&$ 1.148\quad$&$ 1.119\quad$&$ 0.720\quad$&$ 1.116\quad$&$ 1.093\quad$&$
 0.714\quad$  \\
\end{tabular}
\caption{One-loop wave function renormalization for action ${\cal S}^B$ 
in Landau gauge.
Numerical integration errors are as in Table IX.
}
\end{center}
\end{table}
\begin{table}
\begin{center}
\begin{tabular}{c|rrr|rrr}
 \multicolumn{7}{c}{ Action $\act^D \qquad \alpha_g = 1.0$} \\\hline
& \multicolumn{1}{c|}
{$Z_{2,dp_0}^{(1)}$ no $t.i.$ }& 
\multicolumn{1}{c|}
{$Z_{2,dp_0}^{(1)}$ with $t.i.$} &
\multicolumn{1}{c|}
{ $Z_2^{(1)}$ with $t.i.$} &
\multicolumn{1}{c|}
{$Z_{2,dp_0}^{(1)}$ no $t.i.$ }& 
\multicolumn{1}{c|}
{$Z_{2,dp_0}^{(1)}$ with $t.i.$ }&
\multicolumn{1}{c}
{ $Z_2^{(1)}$ with $t.i.$}
\\\hline
$a_sM_1^{(0)}$ & 
\multicolumn{3}{c|}{$\chi = 1.0 \qquad$} &
\multicolumn{3}{c}{$\chi = 2.0 \qquad$} \\\hline
$0.00$&$ 0.619\quad$&$-0.194\quad$&$-0.194\quad$&$ 0.102\quad$&$-0.111\quad$&$
-0.111\quad$  \\
$0.01$&$-2.755\quad$&$-3.568\quad$&$-3.597\quad$&$-3.270\quad$&$-3.482\quad$&$
-3.511\quad$ \\
$0.05$&$-1.636\quad$&$-2.448\quad$&$-2.601\quad$&$-2.146\quad$&$-2.358\quad$&$
-2.510\quad$  \\
$0.10$&$-1.107\quad$&$-1.920\quad$&$-2.185\quad$&$-1.617\quad$&$-1.830\quad$&$
-2.088\quad$  \\
$0.50$&$ 0.320\quad$&$-0.493\quad$&$-1.262\quad$&$-0.287\quad$&$-0.500\quad$&$
-1.146\quad$  \\
$1.00$&$ 1.034\quad$&$ 0.221\quad$&$-0.781\quad$&$ 0.258\quad$&$ 0.045\quad$&$
-0.728\quad$  \\
$2.00$&$ 1.710\quad$&$ 0.898\quad$&$-0.178\quad$&$ 0.735\quad$&$ 0.523\quad$&$
-0.272\quad$  \\
$5.00$&$ 2.138\quad$&$ 1.325\quad$&$ 0.304\quad$&$ 1.217\quad$&$ 1.005\quad$&$
 0.274\quad$  \\\hline
$a_sM_1^{(0)}$ & 
\multicolumn{3}{c|}{$\chi = 3.0 \qquad$} &
\multicolumn{3}{c}{$\chi = 3.6 \qquad$} \\\hline
$0.00$&$ 0.038\quad$&$-0.054\quad$&$-0.054\quad$&$ 0.027\quad$&$-0.036\quad$&$
-0.036\quad$  \\
$0.01$&$-3.332\quad$&$-3.425\quad$&$-3.454\quad$&$-3.344\quad$&$-3.407\quad$&$
-3.436\quad$ \\
$0.05$&$-2.209\quad$&$-2.302\quad$&$-2.452\quad$&$-2.221\quad$&$-2.284\quad$&$
-2.435\quad$  \\
$0.10$&$-1.682\quad$&$-1.775\quad$&$-2.033\quad$&$-1.694\quad$&$-1.757\quad$&$
-2.015\quad$  \\
$0.50$&$-0.383\quad$&$-0.476\quad$&$-1.098\quad$&$-0.403\quad$&$-0.467\quad$&$
-1.082\quad$  \\
$1.00$&$ 0.112\quad$&$ 0.019\quad$&$-0.692\quad$&$ 0.078\quad$&$ 0.015\quad$&$
-0.679\quad$  \\
$2.00$&$ 0.499\quad$&$ 0.406\quad$&$-0.274\quad$&$ 0.438\quad$&$ 0.375\quad$&$
-0.271\quad$  \\
$5.00$&$ 0.878\quad$&$ 0.786\quad$&$ 0.220\quad$&$ 0.770\quad$&$ 0.707\quad$&$
 0.200\quad$  \\\hline
$a_sM_1^{(0)}$ & 
\multicolumn{3}{c|}{$\chi = 4.0 \qquad$} &
\multicolumn{3}{c}{$\chi = 5.0 \qquad$} \\\hline
$0.00$&$ 0.023\quad$&$-0.028\quad$&$-0.028\quad$&$ 0.016\quad$&$-0.016\quad$&$
-0.016\quad$  \\
$0.01$&$-3.348\quad$&$-3.399\quad$&$-3.428\quad$&$-3.354\quad$&$-3.387\quad$&$
-3.415\quad$ \\
$0.05$&$-2.225\quad$&$-2.276\quad$&$-2.426\quad$&$-2.231\quad$&$-2.264\quad$&$
-2.414\quad$  \\
$0.10$&$-1.698\quad$&$-1.749\quad$&$-2.007\quad$&$-1.705\quad$&$-1.737\quad$&$
-1.995\quad$  \\
$0.50$&$-0.412\quad$&$-0.463\quad$&$-1.075\quad$&$-0.424\quad$&$-0.456\quad$&$
-1.063\quad$  \\
$1.00$&$ 0.064\quad$&$ 0.013\quad$&$-0.674\quad$&$ 0.043\quad$&$ 0.011\quad$&$
-0.663\quad$  \\
$2.00$&$ 0.412\quad$&$ 0.361\quad$&$-0.269\quad$&$ 0.372\quad$&$ 0.340\quad$&$
-0.265\quad$  \\
$5.00$&$ 0.720\quad$&$ 0.668\quad$&$ 0.191\quad$&$ 0.635\quad$&$ 0.603\quad$&$
 0.176\quad$  \\\hline
$a_sM_1^{(0)}$ & 
\multicolumn{3}{c|}{$\chi = 5.3 \qquad$} &
\multicolumn{3}{c}{$\chi = 6.0 \qquad$} \\\hline
$0.00$&$ 0.015\quad$&$-0.014\quad$&$-0.014\quad$&$ 0.013\quad$&$-0.009\quad$&$
-0.009\quad$  \\
$0.01$&$-3.355\quad$&$-3.384\quad$&$-3.413\quad$&$-3.358\quad$&$-3.380\quad$&$
-3.409\quad$ \\
$0.05$&$-2.233\quad$&$-2.261\quad$&$-2.411\quad$&$-2.235\quad$&$-2.257\quad$&$
-2.407\quad$  \\
$0.10$&$-1.706\quad$&$-1.735\quad$&$-1.992\quad$&$-1.709\quad$&$-1.731\quad$&$
-1.988\quad$  \\
$0.50$&$-0.426\quad$&$-0.455\quad$&$-1.061\quad$&$-0.430\quad$&$-0.452\quad$&$
-1.057\quad$  \\
$1.00$&$ 0.039\quad$&$ 0.010\quad$&$-0.661\quad$&$ 0.031\quad$&$ 0.009\quad$&$
-0.658\quad$  \\
$2.00$&$ 0.364\quad$&$ 0.335\quad$&$-0.264\quad$&$ 0.350\quad$&$ 0.328\quad$&$
-0.262\quad$  \\
$5.00$&$ 0.618\quad$&$ 0.589\quad$&$ 0.173\quad$&$ 0.586\quad$&$ 0.563\quad$&$
 0.168\quad$  \\
\end{tabular}
\caption{One-loop wave function renormalization for action ${\cal S}^D$ 
in Feynman gauge.
Numerical integration errors are as in Table IX. 
}
\end{center}
\end{table}
\begin{table}
\begin{center}
\begin{tabular}{c|rrr|rrr}
 \multicolumn{7}{c}{ Action $\act^D \qquad \alpha_g = 0.0$} \\\hline
& \multicolumn{1}{c|}
{$Z_{2,dp_0}^{(1)}$ no $t.i.$ }& 
\multicolumn{1}{c|}
{$Z_{2,dp_0}^{(1)}$ with $t.i.$} &
\multicolumn{1}{c|}
{ $Z_2^{(1)}$ with $t.i.$} &
\multicolumn{1}{c|}
{$Z_{2,dp_0}^{(1)}$ no $t.i.$ }& 
\multicolumn{1}{c|}
{$Z_{2,dp_0}^{(1)}$ with $t.i.$ }&
\multicolumn{1}{c}
{ $Z_2^{(1)}$ with $t.i.$}
\\\hline
$a_sM_1^{(0)}$ & 
\multicolumn{3}{c|}{$\chi = 1.0 \qquad$} &
\multicolumn{3}{c}{$\chi = 2.0 \qquad$} \\\hline
$0.00$&$ 1.127\quad$&$ 0.314\quad$&$ 0.314\quad$&$ 0.612\quad$&$ 0.399\quad$&$
 0.399\quad$  \\
$0.01$&$-2.246\quad$&$-3.058\quad$&$-3.097\quad$&$-2.758\quad$&$-2.971\quad$&$
-3.007\quad$ \\
$0.05$&$-1.127\quad$&$-1.939\quad$&$-2.095\quad$&$-1.635\quad$&$-1.847\quad$&$
-2.003\quad$  \\
$0.10$&$-0.598\quad$&$-1.410\quad$&$-1.679\quad$&$-1.106\quad$&$-1.319\quad$&$
-1.580\quad$  \\
$0.50$&$ 0.828\quad$&$ 0.016\quad$&$-0.756\quad$&$ 0.223\quad$&$ 0.011\quad$&$
-0.637\quad$  \\
$1.00$&$ 1.542\quad$&$ 0.730\quad$&$-0.274\quad$&$ 0.768\quad$&$ 0.555\quad$&$
-0.219\quad$  \\
$2.00$&$ 2.219\quad$&$ 1.407\quad$&$ 0.330\quad$&$ 1.246\quad$&$ 1.033\quad$&$
 0.238\quad$  \\
$5.00$&$ 2.646\quad$&$ 1.834\quad$&$ 0.811\quad$&$ 1.727\quad$&$ 1.514\quad$&$
 0.784\quad$  \\\hline
$a_sM_1^{(0)}$ & 
\multicolumn{3}{c|}{$\chi = 3.0 \qquad$} &
\multicolumn{3}{c}{$\chi = 3.6 \qquad$} \\\hline
$0.00$&$ 0.537\quad$&$ 0.444\quad$&$ 0.444\quad$&$ 0.521\quad$&$ 0.457\quad$&$
 0.457\quad$  \\
$0.01$&$-2.833\quad$&$-2.926\quad$&$-2.960\quad$&$-2.849\quad$&$-2.913\quad$&$
-2.947\quad$  \\
$0.05$&$-1.710\quad$&$-1.803\quad$&$-1.959\quad$&$-1.727\quad$&$-1.790\quad$&$
-1.946\quad$  \\
$0.10$&$-1.183\quad$&$-1.276\quad$&$-1.537\quad$&$-1.200\quad$&$-1.263\quad$&$
-1.524\quad$  \\
$0.50$&$ 0.116\quad$&$ 0.023\quad$&$-0.600\quad$&$ 0.091\quad$&$ 0.027\quad$&$
-0.590\quad$  \\
$1.00$&$ 0.610\quad$&$ 0.518\quad$&$-0.194\quad$&$ 0.572\quad$&$ 0.509\quad$&$
-0.186\quad$  \\
$2.00$&$ 0.996\quad$&$ 0.904\quad$&$ 0.224\quad$&$ 0.932\quad$&$ 0.868\quad$&$
 0.222\quad$  \\
$5.00$&$ 1.376\quad$&$ 1.283\quad$&$ 0.717\quad$&$ 1.263\quad$&$ 1.200\quad$&$
 0.693\quad$  \\\hline
$a_sM_1^{(0)}$ & 
\multicolumn{3}{c|}{$\chi = 4.0 \qquad$} &
\multicolumn{3}{c}{$\chi = 5.0 \qquad$} \\\hline
$0.00$&$ 0.514\quad$&$ 0.463\quad$&$ 0.463\quad$&$ 0.504\quad$&$ 0.471\quad$&$
 0.471\quad$  \\
$0.01$&$-2.856\quad$&$-2.907\quad$&$-2.941\quad$&$-2.867\quad$&$-2.899\quad$&$
-2.932\quad$ \\
$0.05$&$-1.733\quad$&$-1.785\quad$&$-1.940\quad$&$-1.744\quad$&$-1.776\quad$&$
-1.932\quad$  \\
$0.10$&$-1.207\quad$&$-1.258\quad$&$-1.519\quad$&$-1.217\quad$&$-1.250\quad$&$
-1.510\quad$  \\
$0.50$&$ 0.080\quad$&$ 0.029\quad$&$-0.585\quad$&$ 0.064\quad$&$ 0.031\quad$&$
-0.578\quad$  \\
$1.00$&$ 0.556\quad$&$ 0.505\quad$&$-0.182\quad$&$ 0.531\quad$&$ 0.499\quad$&$
-0.176\quad$  \\
$2.00$&$ 0.903\quad$&$ 0.852\quad$&$ 0.222\quad$&$ 0.859\quad$&$ 0.827\quad$&$
 0.223\quad$  \\
$5.00$&$ 1.210\quad$&$ 1.159\quad$&$ 0.682\quad$&$ 1.122\quad$&$ 1.090\quad$&$
 0.663\quad$  \\\hline
$a_sM_1^{(0)}$ & 
\multicolumn{3}{c|}{$\chi = 5.3 \qquad$} &
\multicolumn{3}{c}{$\chi = 6.0 \qquad$} \\\hline
$0.00$&$ 0.502\quad$&$ 0.473\quad$&$ 0.473\quad$&$ 0.498\quad$&$ 0.476\quad$&$
 0.476\quad$  \\
$0.01$&$-2.868\quad$&$-2.897\quad$&$-2.929\quad$&$-2.872\quad$&$-2.895\quad$&$
-2.926\quad$ \\
$0.05$&$-1.746\quad$&$-1.775\quad$&$-1.930\quad$&$-1.749\quad$&$-1.772\quad$&$
-1.927\quad$  \\
$0.10$&$-1.219\quad$&$-1.248\quad$&$-1.509\quad$&$-1.223\quad$&$-1.246\quad$&$
-1.506\quad$  \\
$0.50$&$ 0.061\quad$&$ 0.032\quad$&$-0.576\quad$&$ 0.055\quad$&$ 0.033\quad$&$
-0.573\quad$  \\
$1.00$&$ 0.526\quad$&$ 0.497\quad$&$-0.175\quad$&$ 0.517\quad$&$ 0.495\quad$&$
-0.173\quad$  \\
$2.00$&$ 0.851\quad$&$ 0.822\quad$&$ 0.223\quad$&$ 0.835\quad$&$ 0.813\quad$&$
 0.224\quad$  \\
$5.00$&$ 1.104\quad$&$ 1.075\quad$&$ 0.660\quad$&$ 1.071\quad$&$ 1.049\quad$&$
 0.654\quad$  \\
\end{tabular}
\caption{One-loop wave function renormalization for action ${\cal S}^D$ 
in Landau gauge.
Numerical integration errors are as in Table IX.
}
\end{center}
\end{table}
\begin{table}
\begin{center}
\begin{tabular}{c|cccc}
 \multicolumn{5}{c}{ Action $\act^B \qquad \chi = 4.0$} \\\hline
 & regular& tadpole &$C_0^{(1)} \;\;no \;\;t.i.$ (reg + tad) &
$C_0^{(1)} \;\; with \;\; t.i.$ \\\hline
$a_s M_1^{(0)}$& \multicolumn{3}{c} {$\alpha_g = 1.0$}  \\\hline
$0.00$&$-0.618(3)$&$1.474$&$0.856(3)$&$-0.343(3)$\\
$0.01$&$-0.616(8)$&$1.474$&$0.858(8)$&$-0.341(8)$\\
$0.05$&$-0.617(5)$&$1.474$&$0.857(5)$&$-0.342(5)$\\
$0.10$&$-0.617(5)$&$1.474$&$0.857(5)$&$-0.342(5)$\\
$0.50$&$-0.608(3)$&$1.474$&$0.866(3)$&$-0.333(3)$\\
$1.00$&$-0.587(2)$&$1.474$&$0.887(2)$&$-0.312(2)$\\
$2.00$&$-0.547(2)$&$1.474$&$0.927(2)$&$-0.272(2)$\\
$5.00$&$-0.519(2)$&$1.474$&$0.955(2)$&$-0.244(2)$\\
$10.00$&$-0.561(2)$&$1.474$&$0.913(2)$&$-0.286(2)$\\\hline
$a_s M_1^{(0)}$& \multicolumn{3}{c} {$\alpha_g = 0.0$}  \\\hline
$0.00$&$-0.345(3)$&$1.199$&$0.854(3)$&$-0.345(3)$\\
$0.01$&$-0.351(8)$&$1.199$&$0.848(8)$&$-0.351(8)$\\
$0.05$&$-0.347(5)$&$1.199$&$0.852(5)$&$-0.347(5)$\\
$0.10$&$-0.347(5)$&$1.199$&$0.852(5)$&$-0.347(5)$\\
$0.50$&$-0.333(3)$&$1.199$&$0.866(3)$&$-0.333(3)$\\
$1.00$&$-0.312(2)$&$1.199$&$0.887(2)$&$-0.312(2)$\\
$2.00$&$-0.273(2)$&$1.199$&$0.926(2)$&$-0.273(2)$\\
$5.00$&$-0.244(2)$&$1.199$&$0.955(2)$&$-0.244(2)$\\
$10.00$&$-0.286(2)$&$1.199$&$0.913(2)$&$-0.286(2)$\\
\end{tabular}
\caption{One-loop speed of light renormalization for action ${\cal S}^B$ 
for different  $a_sM_1^{(0)}$ values at fixed anisotropy $\chi = 4.0$.
Where errors are not indicated explicitly, they are of $O(1)$ or less 
in the last digit.
}
\end{center}
\end{table}
\begin{table}
\begin{center}
\begin{tabular}{c|cc|cc}
 \multicolumn{5}{c}{ Action $\act^B$} \\\hline
 & $C_0^{(1)} \;\; no \;\;t.i.$& $C_0^{(1)} \;\; with \;\;t.i. \quad$
& $C_0^{(1)} \;\; no \;\; t.i.$ & $C_0^{(1)} \;\; with \;\; t.i. \quad$
 \\\hline
$a_sM_1^{(0)}$ & 
\multicolumn{2}{c|}{$\chi = 1.0 \qquad$} &
\multicolumn{2}{c}{$\chi = 2.0 \qquad$} \\\hline
$0.00$&$-0.039$&$-0.019$&$0.668$&$-0.210$  \\
$0.01$&$-0.036$&$-0.015$&$0.671$&$-0.207$ \\
$0.05$&$-0.038$&$-0.018$&$0.669$&$-0.208$  \\
$0.10$&$-0.044$&$-0.023$&$0.669$&$-0.209$  \\
$0.50$&$-0.127$&$-0.107$&$0.654$&$-0.223$  \\
$1.00$&$-0.265$&$-0.245$&$0.633$&$-0.244$  \\
$2.00$&$-0.509$&$-0.488$&$0.582$&$-0.295$  \\
$5.00$&$-0.749$&$-0.728$&$0.432$&$-0.446$  \\\hline
$a_sM_1^{(0)}$ & 
\multicolumn{2}{c|}{$\chi = 3.0 \qquad$} &
\multicolumn{2}{c}{$\chi = 3.6 \qquad$} \\\hline
$0.00$&$0.807$&$-0.302$&$0.842$&$-0.330$  \\
$0.01$&$0.809$&$-0.300$&$0.843$&$-0.328$ \\
$0.05$&$0.808$&$-0.301$&$0.842$&$-0.329$  \\
$0.10$&$0.807$&$-0.302$&$0.842$&$-0.330$ \\
$0.50$&$0.811$&$-0.298$&$0.849$&$-0.322$  \\
$1.00$&$0.821$&$-0.288$&$0.867$&$-0.304$  \\
$2.00$&$0.836$&$-0.273$&$0.899$&$-0.272$  \\
$5.00$&$0.802$&$-0.307$&$0.907$&$-0.264$  \\\hline
$a_sM_1^{(0)}$ & 
\multicolumn{2}{c|}{$\chi = 4.0 \qquad$} &
\multicolumn{2}{c}{$\chi = 5.0 \qquad$} \\\hline
$0.00$&$0.856$&$-0.343$&$0.880$&$-0.363$  \\
$0.01$&$0.858$&$-0.341$&$0.882$&$-0.361$ \\
$0.05$&$0.857$&$-0.342$&$0.881$&$-0.362$  \\
$0.10$&$0.857$&$-0.342$&$0.881$&$-0.362$  \\
$0.50$&$0.866$&$-0.333$&$0.893$&$-0.350$  \\
$1.00$&$0.887$&$-0.312$&$0.918$&$-0.325$  \\
$2.00$&$0.927$&$-0.272$&$0.969$&$-0.274$  \\
$5.00$&$0.955$&$-0.244$&$1.031$&$-0.212$  \\\hline
$a_sM_1^{(0)}$ & 
\multicolumn{2}{c|}{$\chi = 5.3 \qquad$} &
\multicolumn{2}{c}{$\chi = 6.0 \qquad$} \\\hline
$0.00$&$0.885$&$-0.367$&$0.893$&$-0.374$  \\
$0.01$&$0.887$&$-0.365$&$0.896$&$-0.372$ \\
$0.05$&$0.886$&$-0.366$&$0.895$&$-0.373$  \\
$0.10$&$0.886$&$-0.366$&$0.895$&$-0.373$  \\
$0.50$&$0.898$&$-0.354$&$0.907$&$-0.360$  \\
$1.00$&$0.924$&$-0.327$&$0.935$&$-0.332$  \\
$2.00$&$0.977$&$-0.274$&$0.992$&$-0.275$  \\
$5.00$&$1.046$&$-0.206$&$1.073$&$-0.194$  \\
\end{tabular}
\caption{One-loop speed of light renormalization for action ${\cal S}^B$.
Numerical integration errors are at the $\pm0.008$ 
level for $a_sM_1^{(0)}=0.01$, at the $\pm0.005$ level for $a_sM_1^{(0)}=
0.05$ and $0.10$ and at the $\pm0.003$ level or less for other masses.
}
\end{center}
\end{table}
\begin{table}
\begin{center}
\begin{tabular}{c|rr|rr}
& \multicolumn{2}{c|}{ Action $\act^A \qquad$}
& \multicolumn{2}{c}{ Action $\act^{A^\prime}$} \\\hline
$a_sM_1^{(0)}$ & $C_0^{(1)} \;\; no \;\;t.i.$& $C_0^{(1)} \;\; with \;\;t.i. \quad$
& $C_0^{(1)} \;\; no \;\; t.i.$ & $C_0^{(1)} \;\; with \;\; t.i. \quad$ 
\\\hline
$0.00$&$ 0.000\qquad$&$ 0.000\qquad$&$ 0.000\qquad$&$ 0.000\qquad$  \\
$0.01$&$ 0.000\qquad$&$ 0.000\qquad$&$ 0.000\qquad$&$ 0.000\qquad$ \\
$0.05$&$ 0.000\qquad$&$ 0.000\qquad$&$ 0.000\qquad$&$ 0.000\qquad$  \\
$0.10$&$-0.007\qquad$&$-0.007\qquad$&$-0.007\qquad$&$-0.007\qquad$  \\
$0.50$&$-0.102\qquad$&$-0.102\qquad$&$-0.098\qquad$&$-0.098\qquad$  \\
$1.00$&$-0.258\qquad$&$-0.258\qquad$&$-0.243\qquad$&$-0.243\qquad$  \\
$2.00$&$-0.536\qquad$&$-0.536\qquad$&$-0.490\qquad$&$-0.490\qquad$  \\
$5.00$&$-0.809\qquad$&$-0.809\qquad$&$-0.730\qquad$&$-0.730\qquad$  \\
\end{tabular}
\caption{One-loop speed of light renormalization for actions ${\cal S}^A$ 
and  ${\cal S}^{A^\prime}$.
Numerical integration errors are as in Table XIV.
}
\end{center}
\end{table}
\begin{table}
\begin{center}
\begin{tabular}{c|cc|cc}
 \multicolumn{5}{c}{ Action $\act^D$} \\\hline
 & $C_0^{(1)} \;\; no \;\;t.i.$& $C_0^{(1)} \;\; with \;\;t.i. \quad$
& $C_0^{(1)} \;\; no \;\; t.i.$ & $C_0^{(1)} \;\; with \;\; t.i. \quad$ \\\hline
$a_sM_1^{(0)}$ & 
\multicolumn{2}{c|}{$\chi = 1.0 \qquad$} &
\multicolumn{2}{c}{$\chi = 2.0 \qquad$} \\\hline
$0.00$&$-0.424$&$-0.139$&$\;\;0.213$&$-0.301$  \\
$0.01$&$-0.421$&$-0.137$&$\;\;0.216$&$-0.298$ \\
$0.05$&$-0.424$&$-0.139$&$\;\;0.213$&$-0.301$  \\
$0.10$&$-0.430$&$-0.146$&$\;\;0.211$&$-0.303$  \\
$0.50$&$-0.525$&$-0.241$&$\;\;0.179$&$-0.335$  \\
$1.00$&$-0.683$&$-0.398$&$\;\;0.129$&$-0.385$  \\
$2.00$&$-0.971$&$-0.686$&$\;\;0.033$&$-0.482$  \\
$5.00$&$-1.275$&$-0.991$&$-0.180$&$-0.694$  \\\hline
$a_sM_1^{(0)}$ & 
\multicolumn{2}{c|}{$\chi = 3.0 \qquad$} &
\multicolumn{2}{c}{$\chi = 3.6 \qquad$} \\\hline
$0.00$&$0.320$&$-0.388$&$0.345$&$-0.415$  \\
$0.01$&$0.324$&$-0.385$&$0.348$&$-0.412$ \\
$0.05$&$0.321$&$-0.387$&$0.346$&$-0.414$  \\
$0.10$&$0.319$&$-0.390$&$0.344$&$-0.415$  \\
$0.50$&$0.305$&$-0.404$&$0.334$&$-0.426$  \\
$1.00$&$0.286$&$-0.422$&$0.324$&$-0.436$  \\
$2.00$&$0.256$&$-0.452$&$0.311$&$-0.448$  \\
$5.00$&$0.167$&$-0.541$&$0.266$&$-0.494$  \\\hline
$a_sM_1^{(0)}$ & 
\multicolumn{2}{c|}{$\chi = 4.0 \qquad$} &
\multicolumn{2}{c}{$\chi = 5.0 \qquad$} \\\hline
$0.00$&$0.356$&$-0.427$&$0.372$&$-0.446$  \\
$0.01$&$0.358$&$-0.425$&$0.375$&$-0.443$ \\
$0.05$&$0.356$&$-0.426$&$0.373$&$-0.444$  \\
$0.10$&$0.355$&$-0.427$&$0.372$&$-0.446$  \\
$0.50$&$0.347$&$-0.436$&$0.367$&$-0.451$  \\
$1.00$&$0.340$&$-0.443$&$0.365$&$-0.453$  \\
$2.00$&$0.335$&$-0.447$&$0.372$&$-0.446$  \\
$5.00$&$0.311$&$-0.471$&$0.382$&$-0.436$  \\\hline
$a_sM_1^{(0)}$ & 
\multicolumn{2}{c|}{$\chi = 5.3 \qquad$} &
\multicolumn{2}{c}{$\chi = 6.0 \qquad$} \\\hline
$0.00$&$0.375$&$-0.450$&$0.381$&$-0.457$  \\
$0.01$&$0.378$&$-0.446$&$0.384$&$-0.453$ \\
$0.05$&$0.376$&$-0.448$&$0.383$&$-0.455$  \\
$0.10$&$0.376$&$-0.449$&$0.382$&$-0.455$  \\
$0.50$&$0.371$&$-0.454$&$0.378$&$-0.459$  \\
$1.00$&$0.370$&$-0.455$&$0.379$&$-0.459$  \\
$2.00$&$0.379$&$-0.446$&$0.392$&$-0.446$  \\
$5.00$&$0.396$&$-0.429$&$0.421$&$-0.416$  \\
\end{tabular}
\caption{One-loop speed of light renormalization for action ${\cal S}^D$. 
Numerical integration errors are as in Table XIV.
}
\end{center}
\end{table}

\newpage
\begin{figure}
\begin{center}
\epsfysize=7.in
\centerline{\epsfbox{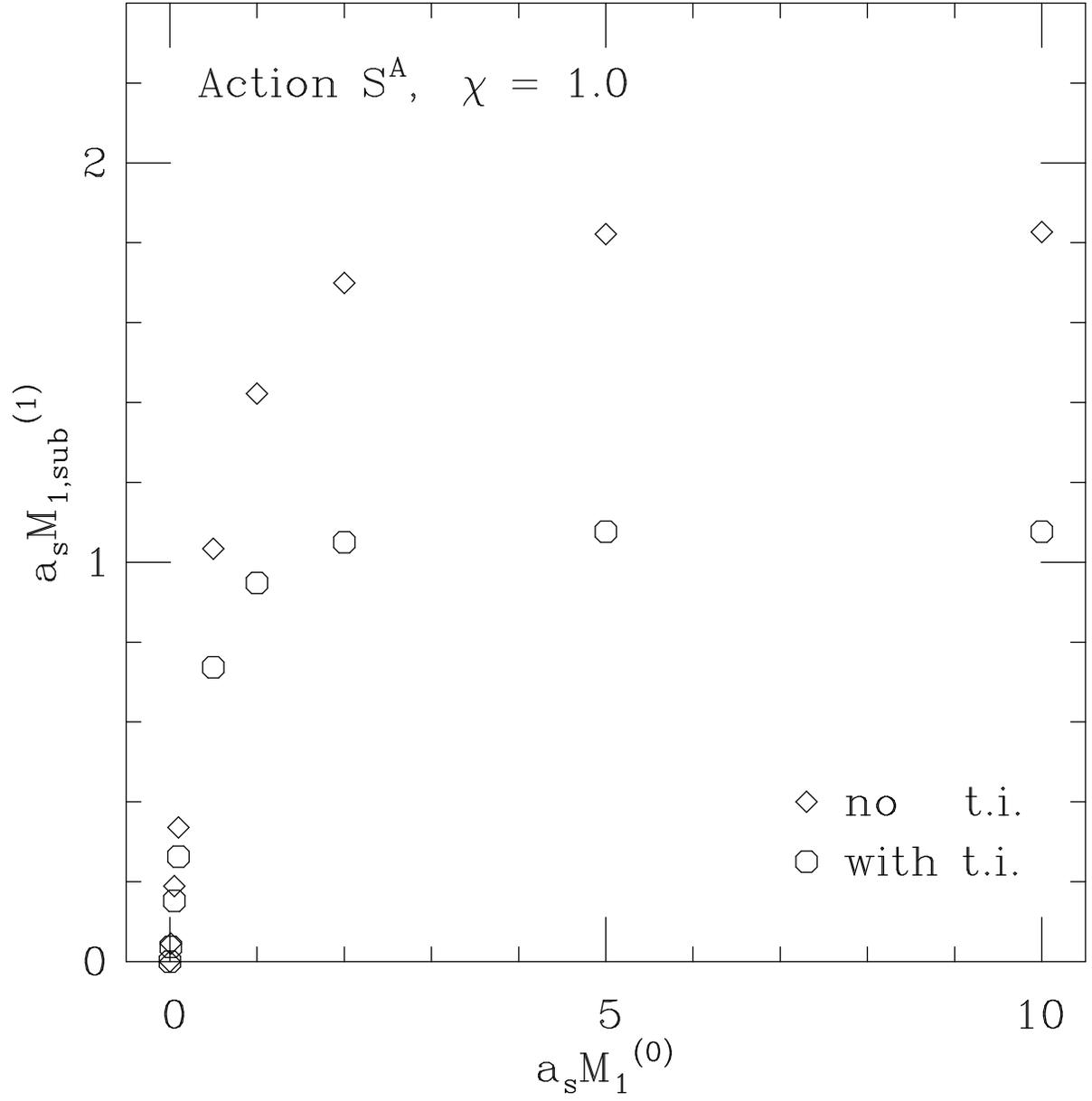 }}
\end{center}
\caption[fig:one]{
  $a_sM_{1,sub}^{(1)}$ versus $a_sM_1^{(0)}$ both with and without tadpole
improvement for action ${\cal S}^A$.
\label{fig:one}}
\end{figure}

\newpage
\begin{figure}
\begin{center}
\epsfysize=7.in
\centerline{\epsfbox{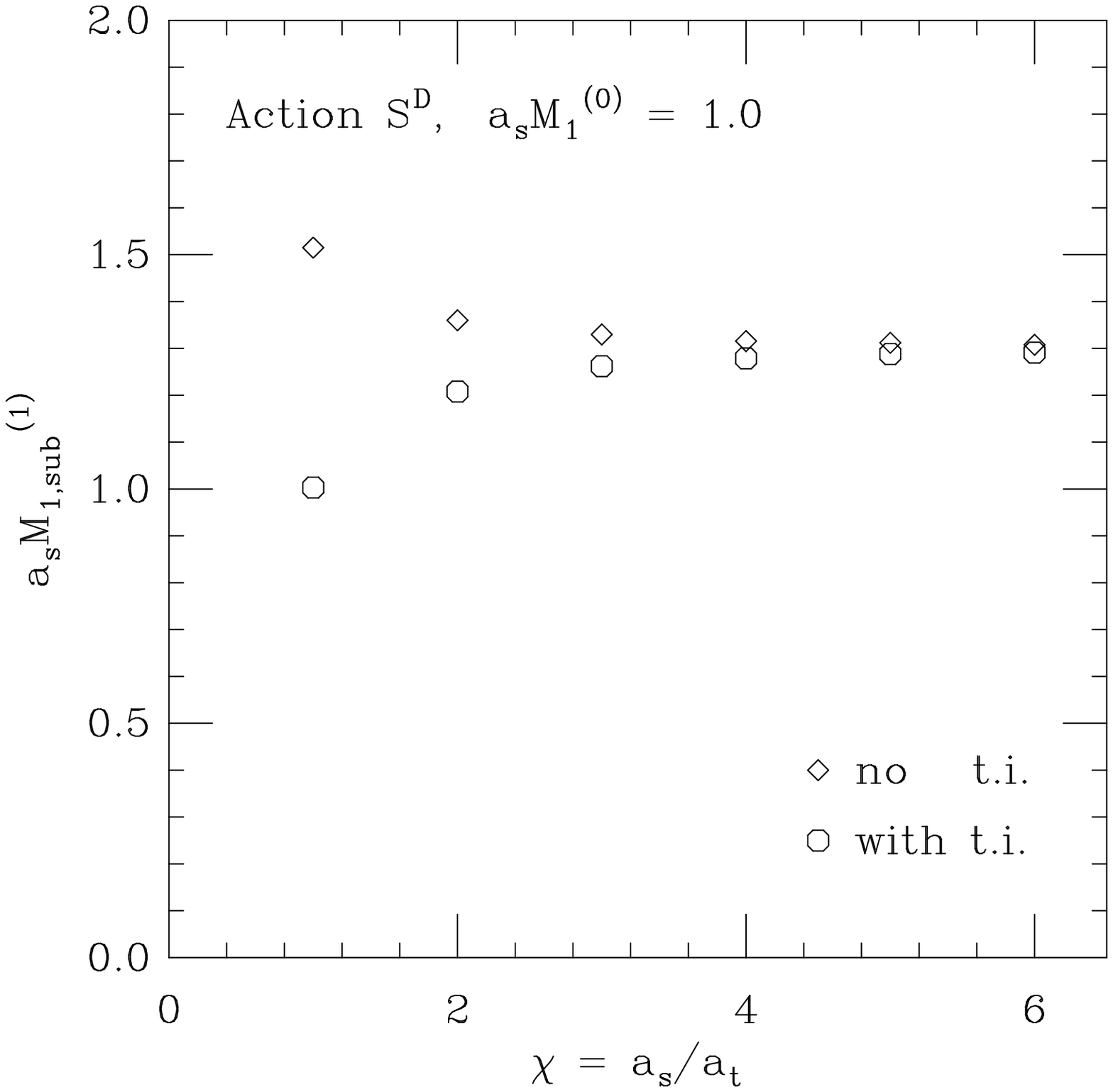 }}
\end{center}
\caption[fig:two]{
  $a_sM_{1,sub}^{(1)}$ versus $\chi$ both with and without tadpole
improvement for action ${\cal S}^A$.
\label{fig:two}}
\end{figure}

\newpage
\begin{figure}
\begin{center}
\epsfysize=7.in
\centerline{\epsfbox{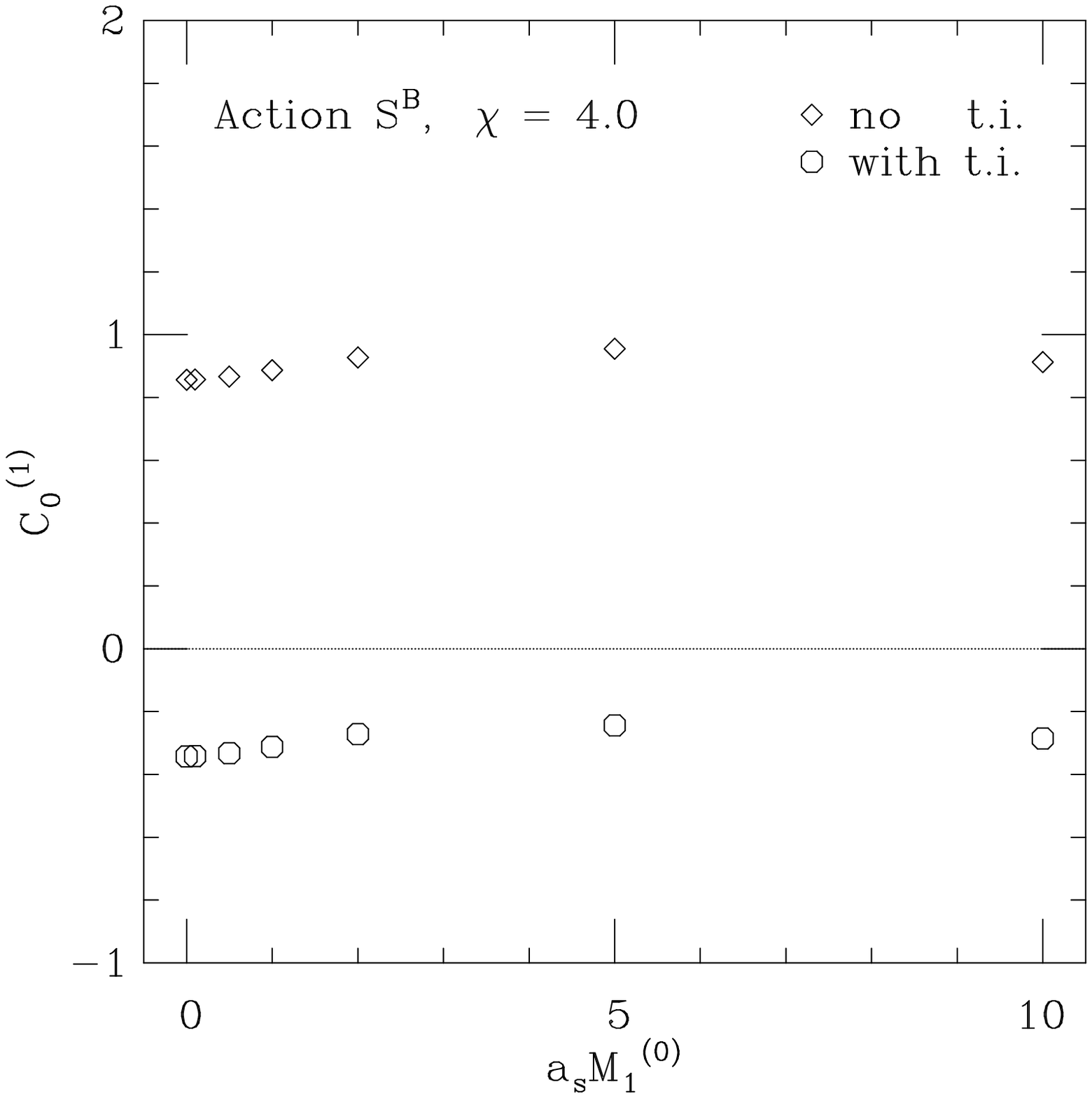 }}
\end{center}
\caption[fig:three]{
  $C_0^{(1)}$ versus $a_sM_1^{(0)}$ both with and without tadpole
improvement for action ${\cal S}^B$.
\label{fig:three}}
\end{figure}

\newpage
\begin{figure}
\begin{center}
\epsfysize=7.in
\centerline{\epsfbox{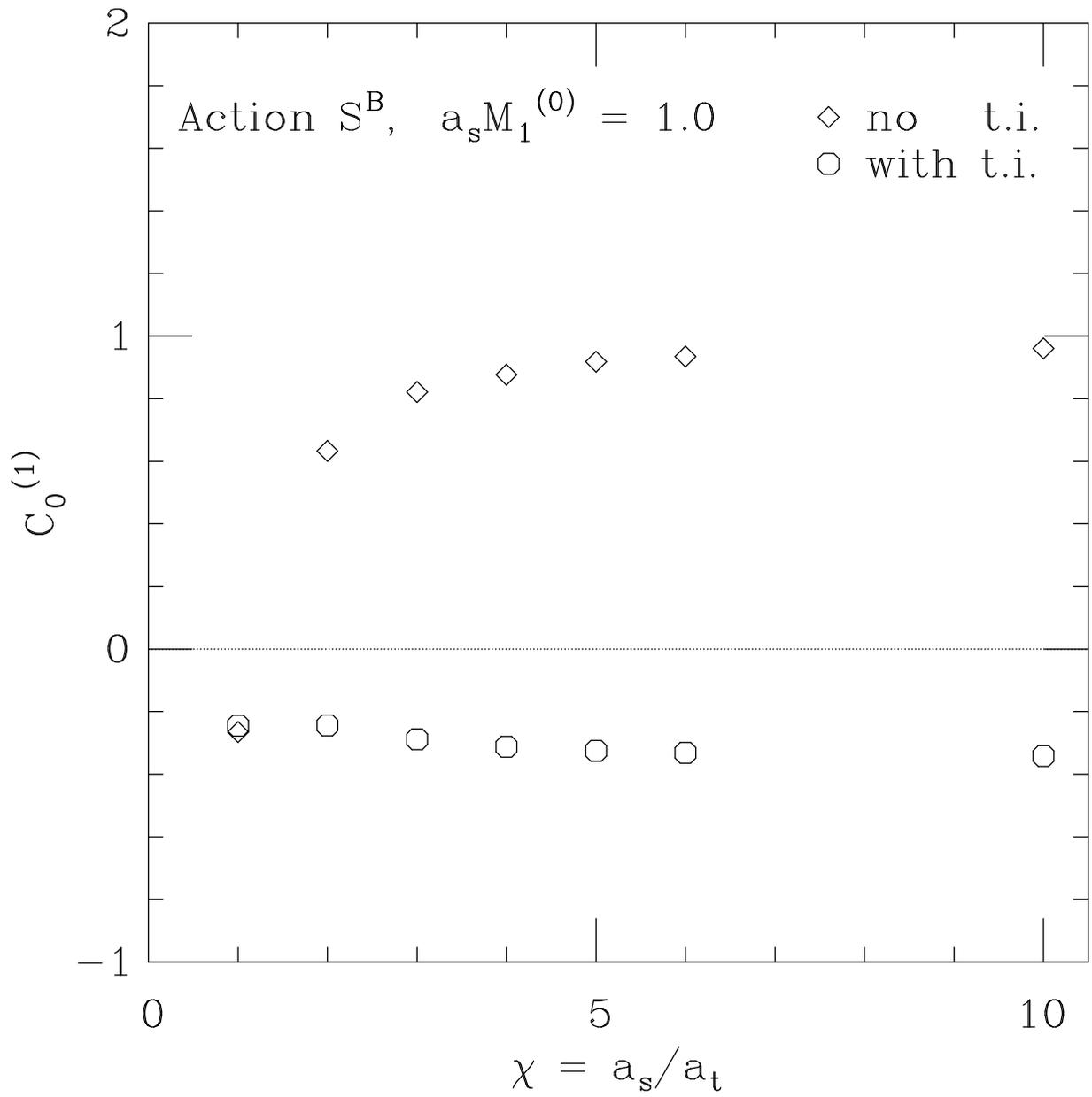 }}
\end{center}
\caption[fig:four]{
  $C_0^{(1)}$ versus $\chi$ both with and without tadpole
improvement for action ${\cal S}^B$.
\label{fig:four}}
\end{figure}

\end{document}